\documentclass[preprint,12pt]{elsarticle}

\usepackage{amssymb}
\usepackage{graphicx}
\usepackage{mathptmx}
\usepackage[T1]{fontenc}
\usepackage[english]{babel}
\usepackage[utf8x]{inputenc}
\usepackage{graphicx}
\usepackage{setspace}
\usepackage[mathscr]{euscript}
 \usepackage{amsmath}
\usepackage[margin=0.75in]{geometry}
\usepackage{float}
\usepackage{tabularx}
\newcolumntype{L}{>{\raggedright\arraybackslash}X}
\usepackage{subcaption}
\usepackage{wrapfig}
\usepackage{multirow}
\usepackage{booktabs}
\usepackage{color}
\usepackage{svg}
 \newcolumntype{C}{>{\centering\arraybackslash}p{2em}}
\usepackage[hyphens]{url}
\usepackage{hyperref}
\usepackage{amssymb}

\usepackage{enumitem}
\usepackage{mdframed}
\usepackage{xcolor}
\usepackage{algpseudocode}
\usepackage{algorithm2e}

\journal{IJEPES}

\begin{document}

\begin{frontmatter}



\title{Edge-Based Detection and Localization of Adversarial Oscillatory Load Attacks Orchestrated By Compromised EV Charging Stations}


\author[inst1]{Khaled~ Sarieddine}
\author[inst1]{Mohammad~Ali~Sayed}
\author[inst2]{Sadegh~Torabi}
\author[inst3]{Ribal~Atallah}
\author[inst1]{Chadi~Assi}

\affiliation[inst1]{organization={The Security Research Center, Concordia University},
            city={Montreal},
            state={Quebec},
            country={Canada}}
\affiliation[inst2]{organization={Center for Secure Information Systems, George Mason University},
            city={Fairfax},
            state={Virginia},
            country={United States of America}}
            
\affiliation[inst3]{organization={Hydro-Quebec Research Institute},
            city={Montreal},
            state={Quebec},
            country={Canada}}

\begin{abstract}
 Recent reports indicate that Electric Vehicle Charging Stations (EVCS) are susceptible to remote exploitation through their vulnerable software/cyber components. More importantly, compromised EVCS can be leveraged to perform coordinated oscillatory load attacks against the interconnected power grid, leading to power grid instability, increased operational costs, and power line tripping. In this paper, we investigate an edge-based approach for the detection and localization of coordinated oscillatory load attacks initiated by exploited EV charging stations against the power grid. We rely on the behavioral characteristics of the power grid in the presence of interconnected EVCS while combining cyber and physical layer features to implement deep learning algorithms for the effective detection of oscillatory load attacks at the EVCS. We evaluate the proposed detection approach by building a real-time test bed to synthesize benign and malicious data, which was generated by analyzing real-life EV charging data collected during recent years. The results demonstrate the effectiveness of the implemented approach with the Convolutional Long-Short Term Memory model producing optimal classification accuracy (99.4\%). Moreover, our analysis results shed light on the impact of such detection mechanisms towards building resiliency into different levels of the EV charging ecosystem while allowing power grid operators to localize attacks and take further mitigation measures. Specifically, we managed to decentralize the detection mechanism of oscillatory load attacks and create an effective alternative for operator-centric mechanisms to mitigate multi-operator and MitM oscillatory load attacks against the power grid. Finally, we leverage the created test bed to evaluate a distributed mitigation technique, which can be deployed on public/private charging stations to average out the impact of oscillatory load attacks while allowing the power system to recover smoothly \textcolor{black}{within 1 second} with minimal overhead.

\end{abstract}


\end{frontmatter}


\section{Introduction}
\label{sec:introduction}
The increasing demands for Electric Vehicles (EVs) in recent years have been driven by a number of factors such as governmental policies and incentives, environmental concerns, rising gas prices, \textcolor{black}{ and a decrease in the prices of EV batteries}, to name a few~\cite{regan_2020,flannery_2021,shingler_2020,gyulai_2020,riley_2021, khan2022multi}. As a result of the rapid adoption of EVs, several public/private entities have invested heavily to accelerate the deployment of the supporting EV Charging Stations (EVCSs) in major cities.
For instance, the Government of Canada has already invested over \$1 billion to support the increased zero-emission EV adoption, with a \$680 million initiative towards addressing the lack of charging and refueling stations in Canada by 2027~\cite{canada_2021}. 
Moreover, EVCSs have been equipped with remote connection and communication capabilities, which \textcolor{black}{facilitate smart charging and scheduling of sessions by the EV users' and the operators' remote managing capabilities of the infrastructure}.

Despite \textcolor{black}{these} benefits, the remote control/management capabilities instilled on the Internet-enabled EVCSs open doors for exploiting the EV charging infrastructure through various vulnerable components in the cyber-layer. \textcolor{black}{The EV ecosystem cyber layer consists of a complex system of interconnected components such as the mobile application, EVCS firmware, the back-end cloud management system (CMS), and the communication links/protocols~\cite{sayed2021electric, alcaraz2017ocpp, rubio2018addressing} to name a few}. In fact, recent reports indicate that the EVCS ecosystem is vulnerable to remote cyber attacks, which have real-life impacts. \textcolor{black}{For instance, in 2022 2 different attacks have been confirmed on the EV ecosystem. In March of 2022, EV charging stations in Russia were hacked and used to display anti-war messages~\cite{independent_2022} while rendering them unavailable to consumers. In the UK as well, EVCSs were hacked and rendered unavailable while inappropriate content was displayed on their screens \cite{mailonline_2022}}. More importantly, while these reported incidents demonstrate the insecurity of the EVCS ecosystem, it raises concerns about the possibility of leveraging such vulnerable devices/systems as an entry point to attack the operations of the interconnected critical infrastructure such as the power gird~\cite{sayed2021electric}. For instance, malware-infected workstations and Supervisory Control and Data Acquisition (SCADA) systems were leveraged to attack the Ukrainian power grid in 2017~\cite{polityuk_vukmanovic_jewkes_2017, perez_2016}, leading to power line tripping and depriving about a quarter million consumers of using electricity for up to 6 hours.

In general, an adversary can exploit the vulnerable components within the EV charging ecosystem (e.g., CMS or the communication links) to create a botnet of infected EVCSs, which can be remotely controlled to launch detrimental attacks against the power grid. For instance, the adversary can command a swarm of EVCSs to start simultaneous charging operations and then stop them repeatedly to impact the stability of the power grid by altering the generator's speed, tripping power lines, overloading the grid, and/or creating frequency instability \cite{sayed2021electric, soltan2018blackiot}. In addition to the possible physical impact, such attacks can also impact the power grid's efficiency and operational costs (e.g., line losses and generation costs) \cite{sarieddine2022investigating, soltan2018blackiot}.
Considering the costly physical and financial implications of large-scale cyber attacks against the power grid, there is a need for effective detection mechanisms on different levels of the multi-layered heterogeneous ecosystem \cite{akhras2020securing} to improve grid resiliency and ensure fault-tolerance. 
%

Previous work discussed various ways to detect attacks and malicious activities on the grid. For instance, in \cite{upadhyay2021intrusion, upadhyay2020gradient, li2021intrusion}, the authors proposed deep learning algorithms to create intrusion detection models based on readings obtained from various input sources such as the Phasor Measurement Unit (PMU), control panel logs, snort network alerts, and relay logs. While these mechanisms mainly depend on monitoring the power grid performance and state estimation to detect abnormal behaviors, they may fail to detect covert and stealthy attacks that are hidden or rendered as normal~\cite{kabir2021two}. Moreover, Kabir et al.~\cite{kabir2021two} devised a cyber-layer detection and mitigation mechanism, which is deployed on the CMS \cite{kabir2021two}. Nevertheless, such centralized detection techniques can be evaded due to the existing vulnerabilities found within the EV charging ecosystem. For instance, adversaries can exploit vulnerable EV's CMS~\cite{sayed2021electric} and/or perform man-in-the-middle (MitM) attacks on the communication protocols (e.g., OCPP~\cite{alcaraz2017ocpp}) to compromise the integrity of the data and the deployed detection models. \textcolor{black}{Such a centralized mechanism in this multilayered and complex ecosystem, provides a single point of failure for the detection and mitigation mechanisms since attackers can easily evade them by attacking other components or layers of the ecosystem}.

\textcolor{black}{In fact, the utility does not have the flexibility to monitor individual consumer loads in real-time, which hinders its ability to detect and localize attacks initiated by the EV charging system and forces it to depend on historical data and grid measurements \cite{soltan2018blackiot}. While an oscillatory load attack might be detected through the physical layer detection if it was not built stealthily, the attack can only be detected through its impact, by which time the damage might have already occurred. Even when the utility is able to locate the bus that is showing abnormal behavior, it would not be able to identify the exact consumer load that was used to alter the grid behavior. This is especially true for the EVCS load which is privately owned (by companies operating the infrastructure) hindering the utility's observability over the ecosystem. It is worth highlighting that even when the utility has some observability over the ecosystem, the utility faces a limitation in localizing the source of the attack on the power grid reaching the granularity of identifying the individual EVCS (location, operator, etc.).This is further amplified by the wide distribution of the EVCS ecosystem and the lack of standardization in the deployment (e.g., multiple operators) which increases the complexity of detecting such attacks and localizing them. Additionally, we highlight that CMS detection mechanisms (operator-centric) are limited to attacks launched by public EVCSs and cannot protect the grid against attacks initiated by private charging stations \cite{kabir2021two}}. Furthermore, while the proposed detection mechanism by Kabir et al.~\cite{kabir2021two} seems to be effective, it only focused on a specific type of oscillatory load attacks. Moreover, it resulted in a high false-negative rate (about $30$\%) for prompt attacks with 20 seconds duration. Finally, their centralized mitigation mechanism \textcolor{black}{does not satisfy the fault tolerance requirements needed to ensure a secure operation of the power grid} due to the deployment of the detection and mitigation mechanisms on the CMS.


We identify a special kind of adversarial attacks which could be initiated against such centralized and operator-centric mechanisms such as:

\begin{itemize}
    \item Multi-operator oscillatory attack, where an attacker can initiate a multi-operator (exploiting multiple EVCSs belonging to different operators) stealthy attacks that might not be detected or mitigated by the deployed mechanisms suggested by Kabir et al. \cite{kabir2021two} since it does not have a holistic view of the different silo-ed operators. The attacker in this case depends on the collective impact of the compromised EVCSs that belong to different operators to harm the underlying infrastructure (e.g. power grid).
    \item Using slow oscillatory attacks, the adversary can split their oscillatory attack load into numerous charging station groups to remain stealthy. For example, the attacker can divide the oscillatory attack period in half by launching the attack on double the number of EVCSs. This allows the adversary to initiate a slow-oscillatory behavior on the individual charging stations to remain stealthier. The attack can be designed in such a way that the aggregate load seen by the grid is the same, however, the behavior of the individual charging station or management system is inconspicuous.
\end{itemize}  

Therefore, \textcolor{black}{due to the failure of centralized detection mechanisms to detect coordinated attacks and the failure of physical layer detection mechanisms to identify and localize the attacks with high granularity, it is of paramount importance to devise a detection and mitigation mechanism that addresses these gaps. To the best of our knowledge, the attack vectors mentioned above were not considered as part of the previous studies which renders their detection mechanisms ineffective based on the source of the attack}. Thus, there is a need to design a detection mechanism that detects attacks initiated by the EV ecosystem against the power grid, which can also address multi-operator load attacks, \textcolor{black}{slow oscillatory stealthy attacks, and all oscillatory load attacks initiated at any vulnerable point of entry into the ecosystem}. 

In this paper, we investigate and propose an edge (EVCS) based deep learning detection mechanism, that can detect oscillatory load attacks in a decentralized manner while providing the operator with an insight into localizing the source of the attack and attributing it to the EVCS ecosystem with the granularity of an individual charging station. We instill resiliency and fault tolerance into the system by distributing the decision-making. We used a unique set of features to extract and mine the behavioral characteristics of a charging station and its connected infrastructure and achieved a 99.4\% accuracy to swiftly detect oscillatory load attack \textcolor{black}{by viewing the first} 5 seconds \textcolor{black}{of the attack whether it is launched by public or private charging stations}. Moreover, we discuss and evaluate a post-detection distributed mitigation mechanism that attenuates the grid oscillation and helps the system recover smoothly. The proposed cyber-layer mitigation mechanism reduces the impact on the grid by distributing the oscillatory load over a period of time while preserving the quality of service for the customers (non-malicious requests that got classified as malicious). After detecting malicious behavior, a random delay block of up to 4 seconds on new requests is added to deprive the adversary of the luxury of coordinating an attack and thus diminish its impact of the attack on the grid. This random block is added to each charging station independently which allows a distributed and lightweight mitigation technique to effectively diminish the attack impact almost instantaneously.

To this end, we summarize our contributions as follows:
\begin{itemize}[leftmargin=*]
    \item This work is among the first to propose a practical and effective mechanism for detecting coordinated oscillatory load attacks on the grid using a botnet of EVCSs, which represents a new attack surface that has been shown to be vulnerable at scale. This work sheds light on the need for effective detection mechanisms at various layers of the EV infrastructure to protect the grid by building resiliency and fault tolerance into it.
    
    \item This work addresses the challenges in collecting EV charging data by exploring real datasets and using our observations to characterize various benign/malicious behaviors and identify main features. We also created a testbed to synthesize a realistic dataset that represents different benign and malicious and studies the impact of the electric vehicle ecosystem.
    
    \item We propose and evaluate an AI-enabled detection mechanism that can be deployed on private and public charging stations. The results indicate the accuracy and effectiveness of the approach with high precision. We tailored deep learning models to attribute suspicious behaviors and achieved about 99.4\% accuracy while allowing for distributed and independent decision-making by integrating hybrid (cyber and physical) features into the decision process. Along with that, the deployment location of our mechanism allowed us to mitigate adversarial attacks (e.g., multi-operator attacks and MitM OCPP attacks) that could be launched against centralized detection mechanisms. \textcolor{black}{Along with that, our approach enables the utility to detect attacks launched from private charging stations as well.}
    
    \item We propose and evaluated an automated distributed lightweight-mitigation mechanism that can effectively neutralize the oscillatory attacks on the power grid. We discuss the practical effectiveness of the method even when possible evasion techniques are employed by adversaries, which makes it a robust and feasible approach. We also take into consideration maintaining a high quality of service for customers, in case of a false positive, by minimizing the delay to a maximum of 4 seconds.
    
\end{itemize}

The remainder of this paper is organized as follows. In Section \ref{sec:background}, we present background information and basic concepts related to the EV ecosystem and related work. In Section \ref{sec:system_model}, we discuss the system model and discuss the methodology and details our proposed detection and mitigation mechanisms. In Section \ref{sec:results}, we detail the experimental results of the distributed detection and mitigation mechanisms. Finally, we evaluate and provide a discussion of the results of our proposed approach in Section \ref{sec:results2} before providing a concluding remark in Section \ref{sec:conclusion}.

\section{Background and Threat Model}
\label{sec:background}
The EV charging ecosystem represents a cyber-physical system, which is composed of interacting hardware and software components.

\subsection{System Overview}
\begin{figure}[t]
	\centering
	\includegraphics[width=0.9\linewidth]{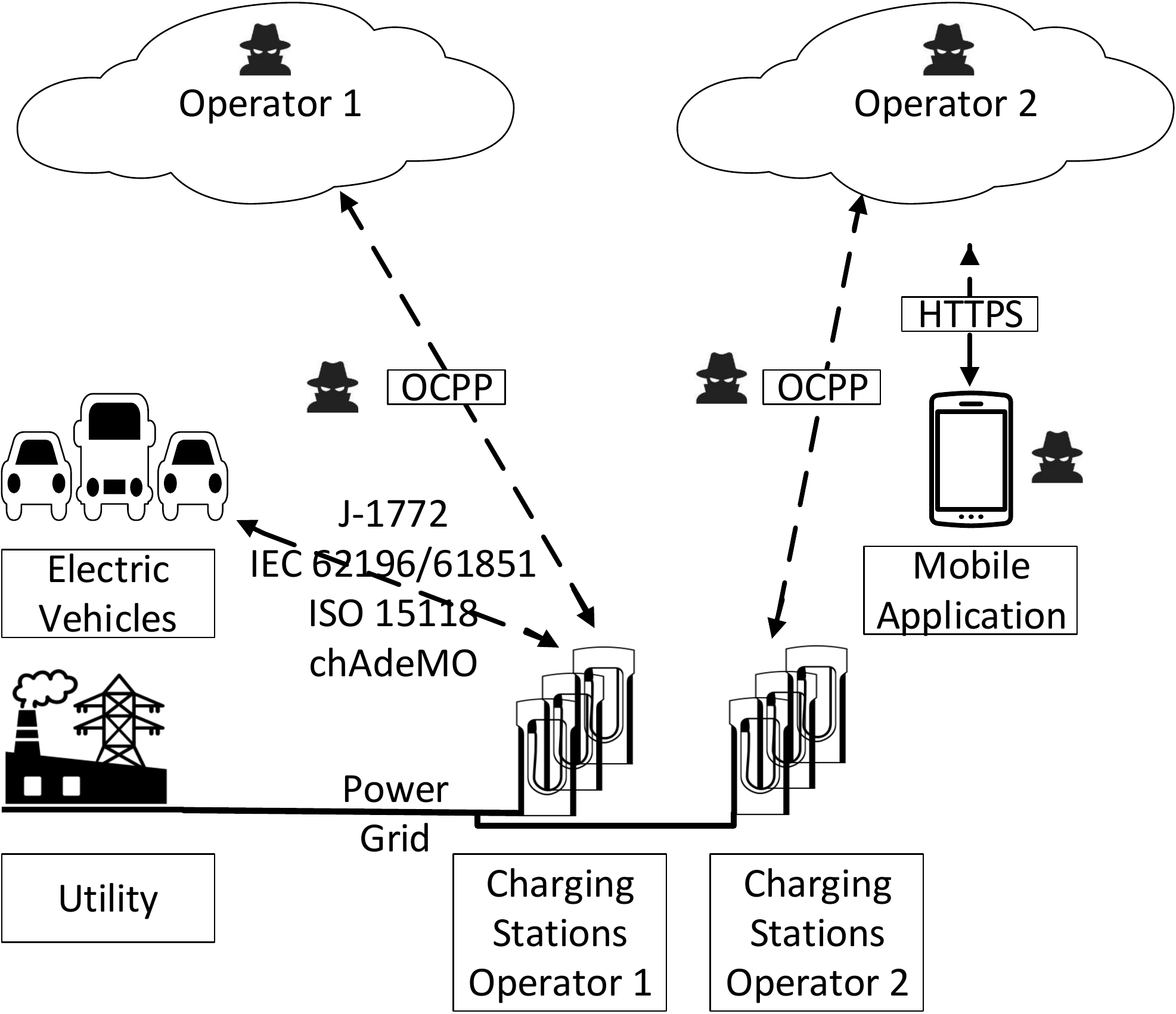}
	\caption{Overview of the EV charging ecosystem and its interactions.}
	\label{fig:ecosystemOverview}
\end{figure}

The software component is composed of the central management system (CMS) and a mobile application. The software component allows remote monitoring and management of the physical counterpart which constitutes the electric vehicle charging station and the vehicle connected to the power grid. The mobile application allows users to control the charging stations remotely and monitor them by sending requests to the CMS. The CMS enables remote communication between the mobile application and the charging station by interpreting commands and controlling the EVCS. The CMS uses the Open Charge Point Protocol (OCPP) \cite{ocpp} which allows it to perform numerous functionalities on the EVCS such as start, stop, update firmware, etc. The OCPP emerged as a result of an effort to standardize the development and deployment of the EVCSs. Moreover, to standardize the communication between the vehicle and the charging station various standards have been put in place such as (SAE J-1772/J-2293/J-2847/J-2836, IEC62196/61851, ISO/IEC 15118, and chAdeMO) \cite{sayed2021electric}. 

There are several types of AC and DC chargers for EVs having different charging rates. Level 2 chargers, which are the most common public charging station types, are being upgraded with faster charging Level 2 or even replaced with new Level 3 chargers (fast/superchargers) to improve the user experience and decrease charging times \cite{sayed2021electric}. These charging stations can either be deployed in public or privately at residences and office buildings. In this work, we focus on both as we aim to detect oscillatory switching attacks originating from private and public charging stations unlike \cite{kabir2021two} that focused on public charging stations. It is worth highlighting that the EVCSs are also connected to the power grid (critical infrastructure) to draw the needed power for vehicles to charge. The connection to the critical infrastructure highlights the importance of securing the deployment of EVCSs against attacks that might impact the power grid.
\subsection{\textcolor{black}{Related Work}}
\label{sec:related_work}
In this section, we survey and discuss previous work that tackled the security of the EV charging ecosystem's components and the attacks that are initiated through the ecosystem against the power grid. Sayed et al. \cite{sayed2021electric}, studied the impact of the oscillatory switching attacks due to several vulnerabilities found in the ecosystem. Moreover, outside of academia Kaspersky Lab's team \cite{kaspersky} analyzed the security of the ChargePoint home charging station and found significant vulnerabilities in its firmware and mobile management application. Moreover, Alcaraz et al. \cite{alcaraz2017ocpp} studied the communication protocol between the management system and the charging stations, which allows the adversary to interfere in the communication between the EVCS and the EV resource reservation service. Whereas, in \cite{sayed2021electric}, the authors studied the impact of oscillatory load attacks initiated by the EVCS against the power grid. The results showed that an insecure ecosystem pauses a great risk to the critical infrastructure to which it is connected. It can lead to system instability, line tripping, etc.

The security vulnerabilities in the EV ecosystem and its critical impact on human lives due to its connection to the power grid \cite{sayed2021electric} motivate the need for a detection mechanism for oscillatory load attacks. In \cite{kabir2021two}, Kabir et al. studied oscillatory load attacks and devised a centralized detection mechanism using a backpropagation neural network. The deep learning model developed can be deployed on a central management system and targets switching attacks that are initiated by the public charging station. Moreover, the proposed approach by Kabir et al. resulted in a 30\% false negative for swift 20 seconds attacks which translates to 30\% of the attacks being classified as normal, the uncertainty in the results motivates the need for an efficient detection mechanism. It is worth mentioning that such operator-centric mechanisms (mechanisms that are deployed on the CMS of each operator) fail to detect multi-operator oscillatory load attacks due to their ability to view the activity of other operators.  In our approach, we proposed a convolutional LSTM model that was able to detect swift attacks with a low false-negative rate. Consequently, using our approach we can defend against multi-operator switching attacks by distributing the decision-making and allowing charging stations to make independent decisions based on shared behavioral characteristics.

Different detection mechanisms have been proposed in the literature to identify physical layer attacks, such as False Data Injection, using recurrent neural networks \cite{ayad2018detection}, or using Bad Data Detection algorithms that rely on measurement residuals \cite{margossian2019partial}. Moreover, other techniques, such as AdaBoost, random forest, and common path mining, have been studied \cite{kosut2011malicious, li2014quickest, wang2021kfrnn, sakhnini2021physical}. The detection of oscillatory load attacks, to the best of our knowledge, hasn't been widely studied in previous work. It is worth mentioning that previous work also lacks localization methods that link attacks to particular physical locations. In our approach because of the portability and our deployment location (EVCS) the operator can identify and localize attacks to the granularity of a charging station which allows the grid operator to create better defenses against attacks. Furthermore, the detection mechanisms that depend on state estimations and knowledge about the power grid using different devices such as PMUs may fail under attacks that target the physical and cyber layers simultaneously \cite{chung2018local}. These types of attacks include steps to mislead the control center similar to the Ukrainian power grid attack.

Moreover, in \cite{basnet2020deep} the authors created an LSTM deep learning model to detect DDoS attacks that can violate the availability of EVCSs by targeting the management system. They studied different types of DDoS attacks that will affect the availability of resources. In our work, we do not assume that the attack has changed the attributes of network packets. Also, we do not have access to network packets before, throughout, and after the attack unlike \cite{basnet2020deep}. We utilize EVCS logs to deploy a distributed detection mechanism on the charging station. Consequently, Basnet et al. furthered their study to create an IDS to detect FDI and DDoS attacks on photovoltaic controllers \cite{basnet2021exploring} whereas we detect oscillatory load attacks on the cyber-layer of the EV ecosystem. Moreover, in \cite{basnet2021ransomware} the authors devised a ransomware detection mechanism while assuming that the ransomware can initiate DDoS and FDI attacks that might alter the state of charge thresholds. The detection mechanism is based on assembly instructions that are generated after the ransomware starts executing, whereas in \cite{sarieddine2021ransomware}, the authors proposed an early detection mechanism based on pre-attack (paranoiac) activity that the ransomware performs before executing. In \cite{basnet2021ransomware}, the authors utilized 561 ransomware samples to train and test their deep-learning model. However, there are various classes/families, wherein \cite{sarieddine2021ransomware} the authors collected about 3000 ransomware samples, which makes the data set created in \cite{basnet2021ransomware} unrepresentative.

\subsection{Oscillatory Attack Vector}
\label{sec:oscillatory}
In \cite{acharya2020public}, the authors exploited publicly available data of EV chargers of the Manhattan, New York, power grid to design a novel data-driven cyberattack strategy using state-feedback-based partial eigenvalue relocation, which targets frequency stability of the power grid. The current number of EVs is not adequate to create sizable impacts, however, with the increased adoption of EVs and deployment of charging stations to match the demand, the grid will face such attacks and impacts. 

To initiate an oscillatory load attack from the EVCS surface, several EVCSs have to alter their charging behavior to follow a repeated on-off behavior within a very short period. The oscillatory attacks are characterized by the EV load, duration of the attack, and the instant of switching. These characteristics differ based on the power grid and its loaded conditions. Two variations of the attack exist, charging oscillatory attack, which relies on starting and stopping several charging stations. Whereas, the discharging oscillatory attack relies on charging and discharging connected EVs through several charging stations.

Different combinations of the two variations can also be included; however, in our work, we focus on the charging oscillatory attacks, whereas future studies could include the discharging paradigm, vehicle-to-grid (V2G), as it gets rolled out to the public. The oscillatory load attack takes advantage of load manipulation and alternates between a surge in demand which causes a frequency drop on the power grid and when the system starts its recovery and the generators start speeding up again the attacker would switch off the EVCS initiated in the first step and cause a frequency increase. This could be amplified by using discharging oscillatory load attacks, which would cause the generators to speed up due to the mismatch between the demand and extra generation \cite{sayed2021electric}.

\textcolor{black}{Different types of oscillatory load attacks can be curated and are summarized as follows:} 
\begin{itemize}
    \item \textcolor{black}{Switching attacks:}
         \begin{itemize}
             \item \textcolor{black}{Square wave: synchronizing the compromised load and switching them between on and off \cite{hammad2017class, sarieddine2022investigating}. This attack can be made stealthier by distributing the switching behavior on multiple EVCSs to reduce the number of events per EVCS}.
             \item \textcolor{black}{Alternating sine wave: synchronizing only small portions of the compromised load every time step \textsc{t} \cite{ghafouri2022coordinated, kabir2021two} (stealthier than square wave attacks and detecting them is not straightforward).}
         \end{itemize}
    \item \textcolor{black}{Dynamic attacks: the size and the trajectory of the compromised load is determined by the attacker based on the grid behavior to achieve and maximize the impact on the grid instability \cite{sayed2022dynamic}.} 
\end{itemize}

\textcolor{black}{From a grid perspective, oscillatory EV loads can be manipulated to have lower power factors \cite{hodge2019vehicle}, thus entailing a larger impact compared to residential loads \cite{sayed2021electric}. Oscillatory load attacks do not require huge loads or injections to cause abnormal behavior on the power grid \cite{sarieddine2022investigating}. Even when the load is not large enough to cause generator tripping, a sustained switching attack can cause frequency and voltage oscillations, which in turn damages the turbines due to the constant acceleration and deceleration \cite{sayed2021electric}. Moreover, it is worth mentioning that a variation of these attacks might target inter-area frequency as discussed in \cite{kabir2021two}, these attacks are stealthy and may not be distinguished from the load variations of the grid \cite{kabir2021two, hammad2017class} which makes oscillatory load attacks initiated by the EVCS ecosystem a serious concern. Furthermore, other oscillatory load attacks can be used to force different types of oscillations, such as exciting sub-synchronous resonance \cite{du2021modeling}. Finally, in the dynamic attack scenario, the adversary induces forced oscillation without the need to excite a specific unstable mode present in the power grid \cite{sayed2022dynamic}.} 

\textcolor{black}{It is worth noting that, the existence of various operators and the wide distribution of the charging stations, created stealthy attack vectors (adversarial) that might exploit the charging stations of different operators to create the same impact on the power grid and hinder the utilities' ability to detect and localize due to the increased complexity in monitoring the consumer loads. Consequently, to locate an attack a utility might depend on PMU measurements and other artifacts however, they do not reach the granularity of identifying the exact location of the charging station that was exploited to initiate the attack due to the wide distribution of the charging stations and the presence of multiple operators. Granular localization information is necessary for the utility to provide adequate countermeasures and create future plans to secure its system.}
\subsection{Threat Model}
We consider an adversary that is able to compromise and control a large number of EVCSs. \textcolor{black}{There are multiple attack vectors which can be used by the adversary to impact the power grid that we take into consideration in our detection mechanism. Namely, we describe the different attack vectors below:}

\begin{itemize}
    \item \textcolor{black}{The internal components of an electric vehicle that have internet connectivity such as the On-Board Diagnostics (OBD) port that can be accessed physically or wirelessly and grant access to the Controller Area Network (CAN) bus, which could be leveraged by the attacker to control the vehicle and its charging \cite{acharya2020cybersecurity}.}
    \item \textcolor{black}{The mobile application which is the component responsible and the enabler for the commercialization of the EVCS ecosystem could be used by the adversary by leveraging the lack of end-end authentication between the user and his vehicle \cite{sarieddine2022investigating}, allowing the adversary to opportunistically take advantage of connected vehicles to the charging station. }
    \item \textcolor{black}{CMSs have been found to be vulnerable to remote attacks. The adversary can exploit one or more operators' management systems (multi-operator) the adversary can perform attacks against the power grid \cite{sayed2021electric} by commanding a large distributed EVCS botnet. The adversary could create different combinations of attacks by leveraging multiple CMSs. }
    \item \textcolor{black}{The OCPP protocol is also taken into consideration which has been found vulnerable to MitM attacks that could be used to initiate and bypass any protection mechanism deployed on the cloud \cite{alcaraz2017ocpp}.}
\end{itemize}

\textcolor{black}{Consequently, we highlight that unlike Kabir et al. \cite{kabir2021two}, the attacker, after gaining control of the EVCSs, can launch various types of oscillatory load attacks not limited to inter-area oscillation oscillatory load attacks. Moreover, the adversary does not necessarily take advantage only of public charging stations but could also leverage privately owned charging stations.}

\begin{figure}[t]
    \centering
    \includegraphics[width=0.9\linewidth]{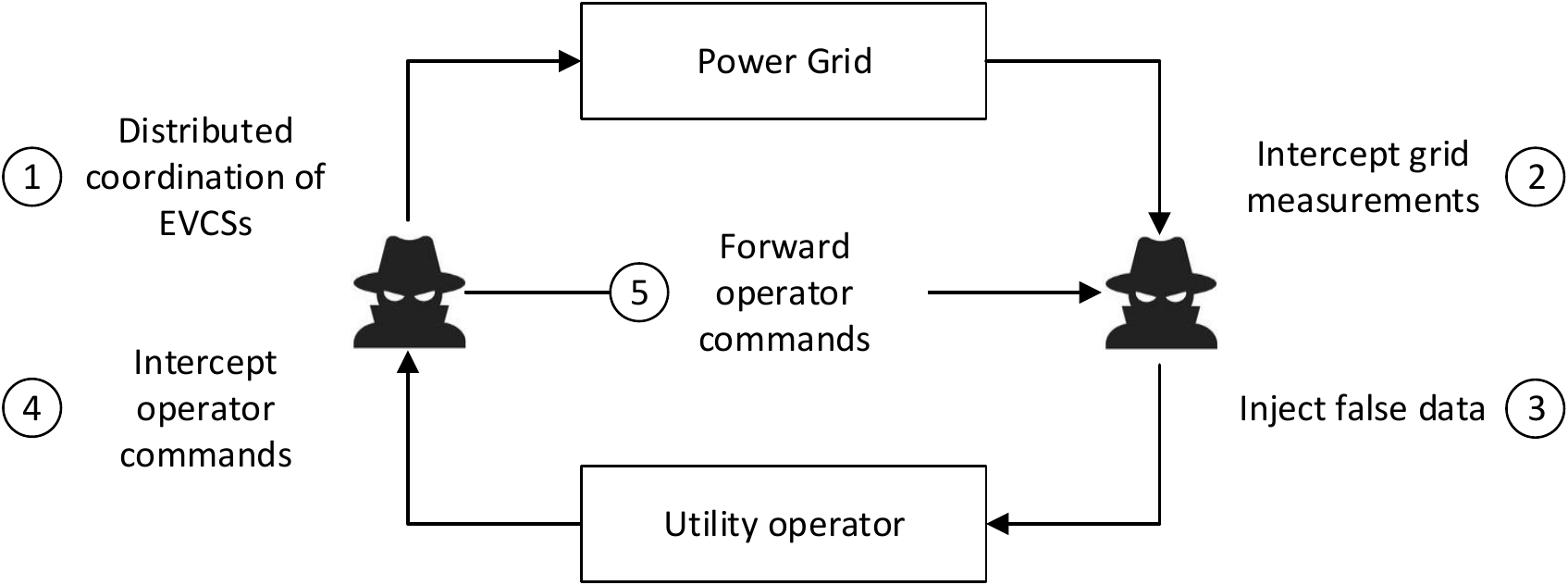}
    \caption{Overview of the covert attack. }
    \label{fig:covert}
\end{figure}

We assume that the attacker can launch covert attacks \cite{lucia2021covert} as illustrated in Figure \ref{fig:covert}. The adversary, as shown in Step 2, controls a considerable number of EVCSs and can command a coordinated oscillatory load attack against the grid. However, to thwart the utility operator's detection mechanisms, the adversary intercepts (Step 2) measurements and readings that the operator collects to monitor and estimate the state of the grid and injects false data (Step 3) which deceives the physical-layer detection mechanism hosted by the utility operator, and thus renders the grid oblivious of the grids' actual state. It is worth mentioning that the adversary injects data that resembles the normal behavior of the grid. Consequently, the utility operator sends commands to the power grid components (e.g., generators) to perform some actions to stabilize the grid based on historical data (e.g., load demand trends), thus the adversary intercepts these commands (Step 4) and forwards them to the false data injector so that the operator can see expected data trends and would not trigger an alarm at the physical layer (Step 5). The attacker can establish covert channels by injecting malware/ransomware \cite{basnet2021ransomware} (e.g., BlackEnergy malware injected into Ukraine's power grid \cite{perez_2016}, Stuxnet Malware infected Iran power grid \cite{maddox_2010}) into the networked controller and arbitrarily alter the control logic. \textcolor{black}{In our work, these threat vectors are addressed using our detection and mitigation mechanisms. Where the attacker's main goal is to induce forced oscillations that would impact the frequency of the grid.}

\textcolor{black}{Now, oscillatory load attacks require the coordination of numerous charging stations simultaneously, and we acknowledge that the current number of EVCSs is not enough to launch the proposed attacks. However, with the current exponential increase in the adoption of electric vehicles and the rapid deployment of EVCSs to match the adoption rate, such attacks pose a great threat to power grid stability. To demonstrate the feasibility of such attacks we chose the New South Wales (NSW) grid whose size is similar to the NE-39 bus grid we use for our dataset collection. The NSW grid has an average load of 6989MW \cite{aemo_2022} and a total number of registered vehicles of 5,892,206 \cite{australianas}. Scaled to fit the 6097MW 39- bus grid, the total number of vehicles in our grid would be 5,155,681. If we assume a future projection of 50\% EV penetration, our grid will contain over 2.5 million EVs. As per the International Energy Agency (IEA) \cite{iea}, based on the mixture of available EVCSs, the average charging rate per EVCS is 24kW. We highlight that based on these statistics, our attacks only require a small portion of the available EVs to be successful. Our largest attack magnitude, for instance, constitutes 30\% of the grid load. This translates to only 3\% of the available EVs. By comparison, our smallest attack magnitude only requires 1\% of the available EVs. When this analysis is performed for the 9-bus system, used in our distributed mitigation section, we see that it only requires 2.6\% of the available EVs if we assume a 50\% penetration level}.

\section{Methodology and System Model}
\label{sec:system_model}
\textcolor{black}{We discuss the methodology and conceptual model of our detection and mitigation. We provide a discussion of the system model, followed by a detailed discussion of our distributed detection methodology. We also discuss the data-set curation and collection. Finally, we discuss our distributed mitigation methodology and provide an overview of the real-time co-simulation testbed on which we demonstrate our mitigation mechanism.}
\subsection{System Model}
\textcolor{black}{In our approach, we attempt to ensure fault tolerance in the deployment of the detection mechanism while handling oscillatory and adversarial oscillatory attacks.} To mitigate covert sophisticated attacks that allow adversaries to deceive traditional physical-layer detection mechanisms, detection should occur at different levels of the interconnected system to build resiliency into it. \textcolor{black}{To address the limitation of previous work, e.g., \cite{kabir2021two} and other centralized detection mechanisms,} we propose to deploy a deep learning model on the EVCS since, the EVCS possesses the ability to collect information about the true operations of the charging stations and power characteristics (e.g., frequency). \textcolor{black}{This presents an advantage over CMS-based detectors} where a compromised OCPP connection allows the adversary to inject bi-directional false data that would affect the detection mechanism deployed there. Moreover, the EVCS is the component that is utilized by adversaries to perform physical attacks by compromising other components (e.g., mobile application, CMS, or OCPP). Thus, securing the EVCSs would prevent attacks initiated from \textcolor{black}{any vulnerable point in the EVCS} ecosystem. Finally, centralizing the detection mechanisms creates a single point of failure, and maximizes the risk of exposing the deep learning model, and polluting the data since recent incident reports and studies show the vulnerability of the system at scale. Whereas the deployment of a deep learning model on the EVCS would hinder the ability of the adversary due to the distributed and independent operation of charging stations (increased resiliency).

\begin{figure}[t]
    \centering
    \includegraphics[width=0.9\linewidth]{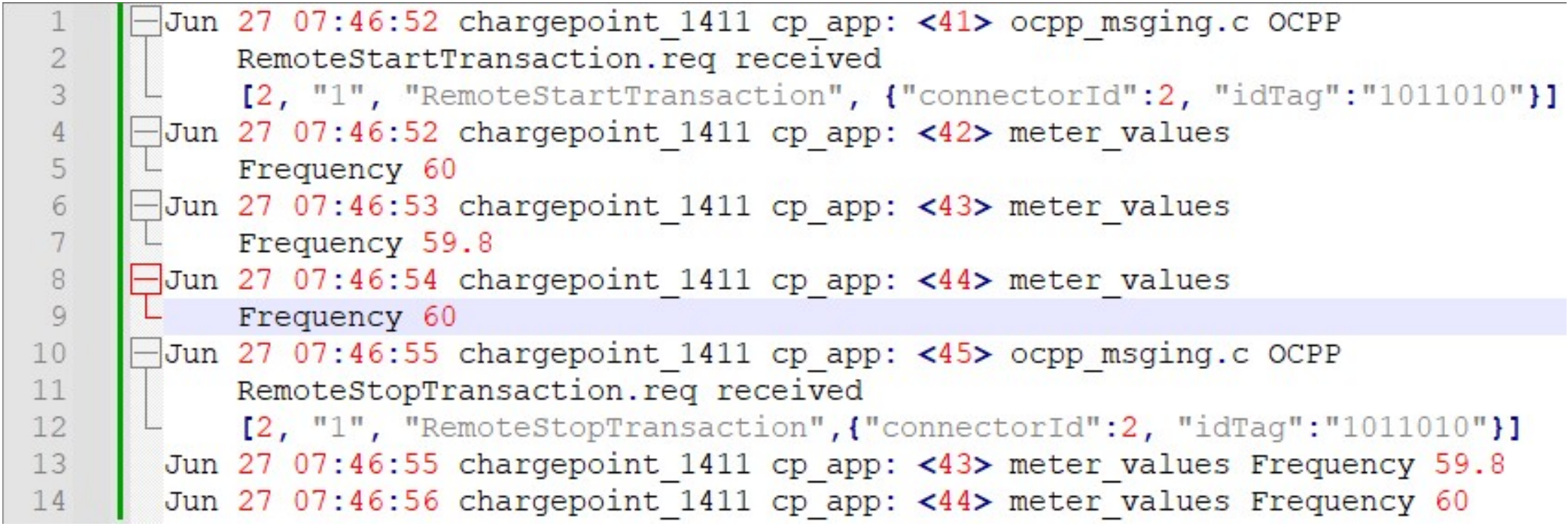}
    \caption{EVCS log showing the different features that could be extracted from the charging station logs.}
    \label{fig:evcslog}
\end{figure}

\begin{figure}[t]
    \centering
    \includegraphics[width=0.9\linewidth]{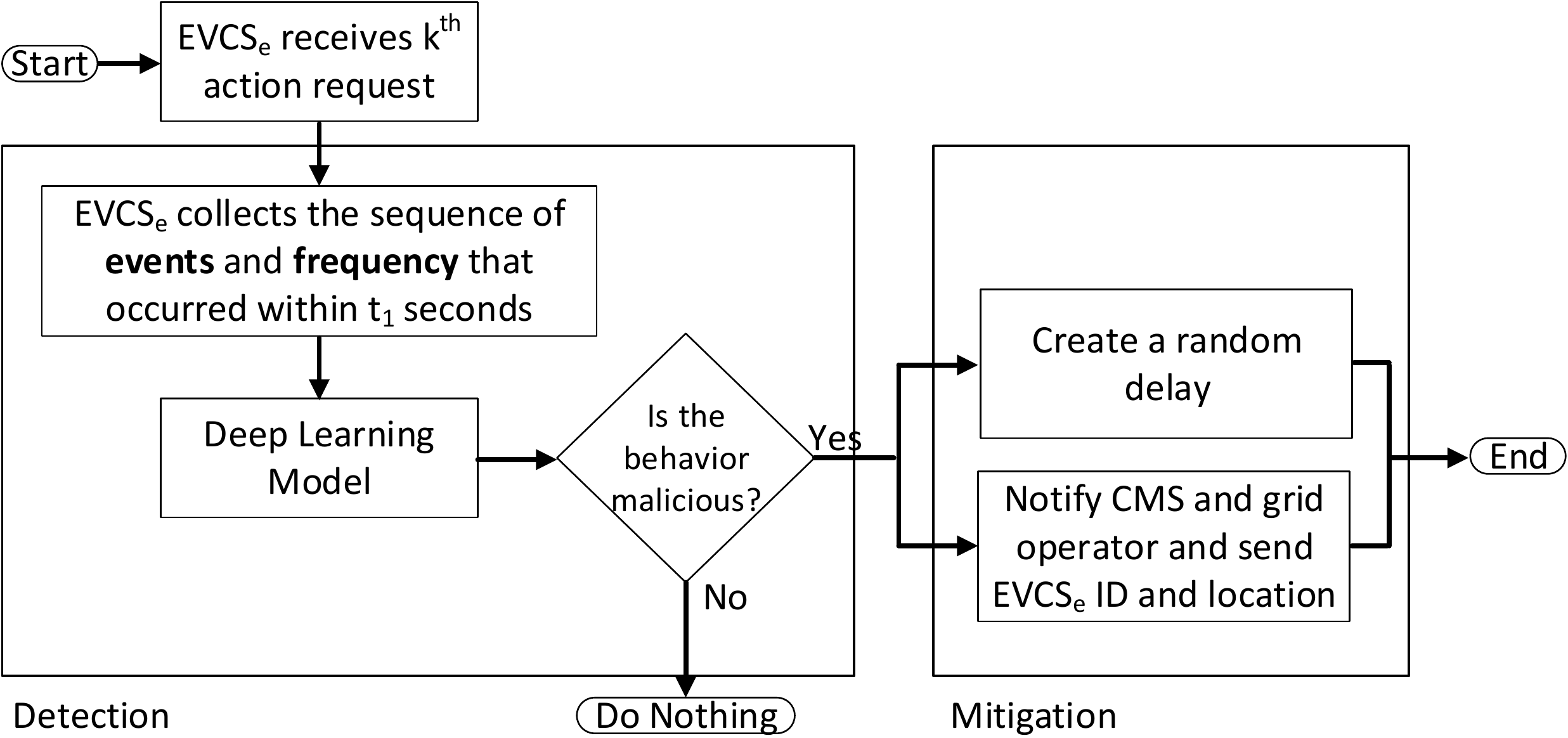}
    \caption{Flow chart describing our detection mechanism.}
    \label{fig:flowChart}
\end{figure}

Consequently, to deploy a deep learning model on the charging station, new features should be derived compared to the work of \cite{kabir2021two}, which utilizes information that only the CMS has access to (e.g., change in a load of vehicles during a certain $\delta$ time). Thus, we investigate the usage of various deep learning techniques in detecting attacks against the power grid initiated by the EVCS ecosystem. To the best of our knowledge, we are among the first to investigate a decentralized cyber-detection mechanism deployed on the EVCS ecosystem to protect the grid from the new vulnerabilities of this cyber-physical system. \textcolor{black}{Residential and public charging stations both have log files} to record all the operations/events of this station. \textcolor{black}{In Figure \ref{fig:evcslog} we show a sample EVCS log where each transaction might include the following information: EVCS ID, operation type (e.g., charging, or stop), operation date, start operation time, stop operation time, charging rate, type of the charger, and the variation of the frequency of the load bus that the EVCS is connected to overtime.} It is worth noting that the OCPP protocol provides a functional block that enables charging stations to send periodic meter values (e.g., voltage, reactive power, etc.). Thus, using the telemetry data collected by the charging station, the power grid frequency, which is directly linked to the speed of the generators, can be \textcolor{black}{directly recorded by the EVCS with high granularity} by measuring the period of the voltage waveforms that are sampled over time. It is worth highlighting that electric devices (e.g., charging stations) will exhibit the same frequency as the bus they are connected to. Thus, to collect grid frequency measurements the utility monitors and collects measurements from the buses which incidentally have connected EVCSs. The recent industrial technology advances increased the connectivity of cyber-physical systems that are monitored and controlled by Supervisory Control and Data Acquisition (SCADA) systems that use advanced computing, sensors, control systems, and communication networks \cite{khoury2020hybrid}. SCADA systems allow power grid operators to gather real-time telemetry data about the grid. This information that the grid operator can acquire from the buses can be used for training deep learning models since, when deploying the model each charging station should be able to gather this information by itself.

Accordingly, the charging station can store each operation in its log file and use it to detect anomalies in the usage of the charging station, which indicates that there is a possible attack initiated from the EVCS ecosystem against the grid. This information can be updated in the log file actively. However, since the utility (power grid operator) is the main entity affected by oscillatory attacks, it will take responsibility to gather information from different operators and distribute trained global models to the connected EVCSs. Collaboration with the utility by various EVCS operators is mandatory to allow a collective view of multi-operator attacks. The utility will use past data to train and deploy deep learning models on the charging stations \textcolor{black}{to alleviate any future privacy concern the operators might have about sharing their data}.

\textsc{Detection Mechanism:} In Figure \ref{fig:flowChart} we give an overview of the proposed detection mechanism to be deployed at the charging station. When an EVCS$_{e}$ receives a charging request (k$^{th}$ request), the EVCS$_{e}$ retrieves the events that occurred in the last t$_{1}$ seconds from its logs. Similarly, it retrieves the frequency readings that has occurred and it collected within the same period from its logs. This information is fed into a machine/deep learning model to detect maliciousness of the events that occurred within the last t$_{1}$ seconds. By leveraging the combination of the cyber data (series of events) and physical data (frequency on the power grid) that are tightly coupled in case of a coordinated oscillatory load attack, we create a deep learning model that will extract the temporal and spatial relationships between the sequence of readings over time. The observed behavior of the charging station and the underlying infrastructure is used to characterize oscillatory load attacks and differentiate them from the normal functioning of a charging station. \textcolor{black}{It is worth highlighting, the t$_{1}$ seconds is a rolling window, and the detection mechanism is real-time. We only require the t$_{1}$ window to detect the attack. This also means that no additional extensive data logs are required to be kept on the EVCS$_{e}$ since our algorithm will not use any of the data prior to the t$_{1}$ rolling window.}

\textsc{Mitigation Mechanism:} If the deep learning model labels the sequence of events as malicious, the EVCS$_{e}$ will create a delay block that will randomly delay request between 0 to 4 seconds to disrupt the synchronization of the oscillatory load attack and notifies the CMS and grid operator by sending the EVCS ID and location. The mitigation mechanism allows distributed and independent decision-making for each charging station, thus ensuring fault tolerance in our mitigation mechanism. Consequently, we test our mitigation mechanism on our test bed which is used to study the impact of the EV ecosystem on the power grid. The results (discussed later) show the effectiveness in neutralizing the impact of an oscillatory load attack on the generator's speed and minimizing the risk and the costs incurred by a successful attack. It is worth highlighting that the independence of our techniques from the features or artifacts that need a global knowledge of the ecosystem and grid provides us with the flexibility needed to deploy our detection-mitigation mechanism on public and private EVCSs.

\subsection{Distributed Detection Mechanism Methodology}
\label{sec:methodology}
Given the limited number of previous works, which discuss the detection and localization of oscillatory load attacks, along with the limitation of previous detection approaches, we aim to deploy an edge-based AI-enabled detection mechanism on the charging station itself. We leverage cyber and physical characteristics (e.g., charging events and power grid frequency variation) to identify and characterize malicious and benign behaviors. \textcolor{black}{More specifically, the devised methodology attempts to leverage the behavioral characteristics of an oscillatory load attack to propose an effective edge-based, decentralized oscillatory load attack detection.}

\begin{figure*}
     \centering
     \begin{subfigure}[b]{0.48\textwidth}
         \centering
         \includegraphics[width=\textwidth]{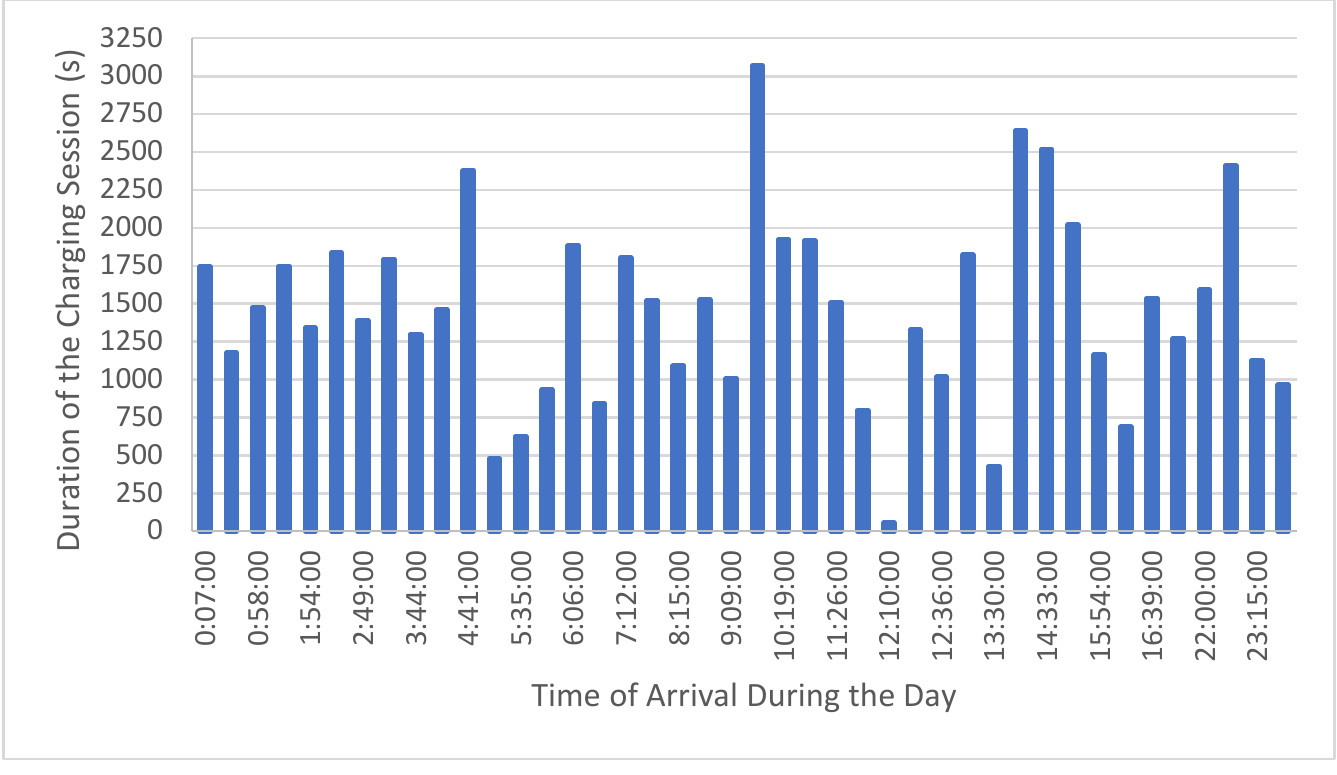}
         \caption{Charging Station 1}
         \label{fig:station1}
     \end{subfigure}
     \hfill
     \begin{subfigure}[b]{0.48\textwidth}
         \centering
         \includegraphics[width=\textwidth]{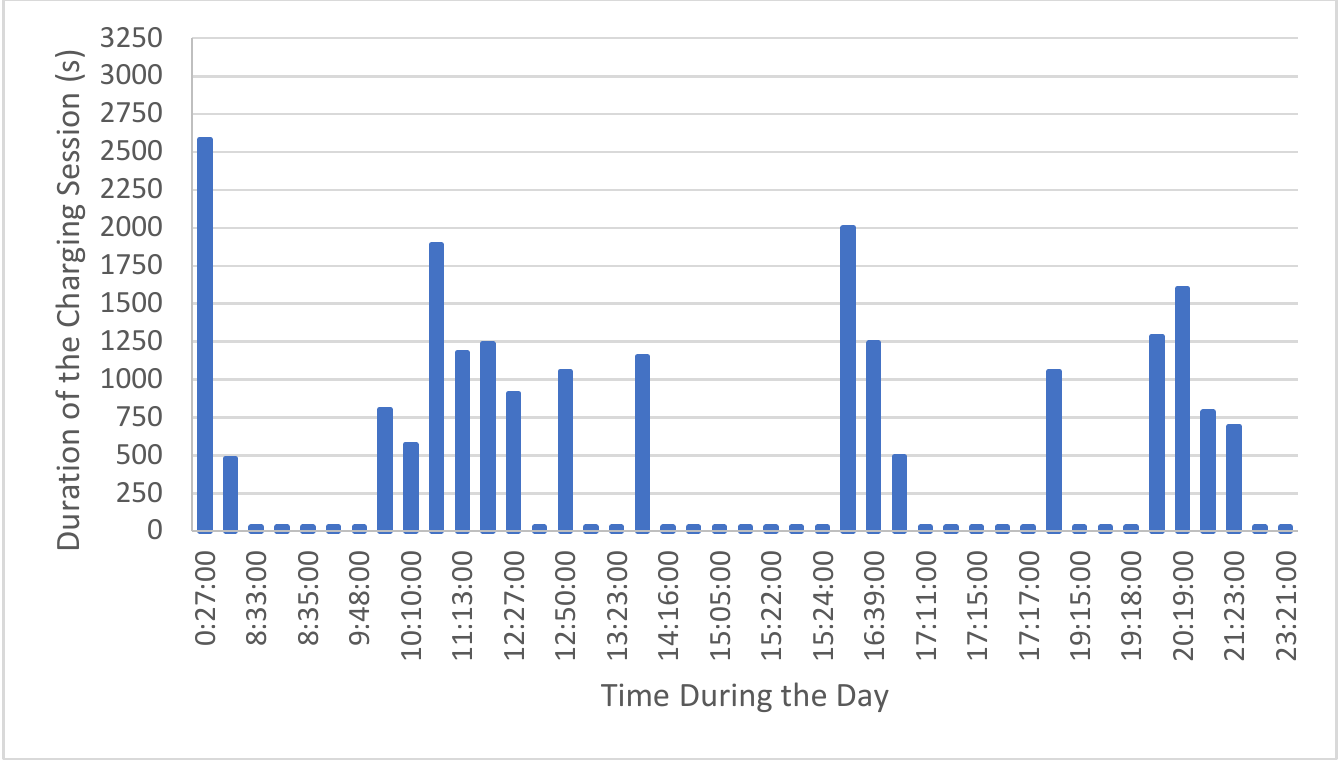}
         \caption{Charging Station 2}
         \label{fig:station2}
     \end{subfigure}
        \caption{Normal charging behavior of two different charging stations.}
      \label{fig:charging behavior}
\end{figure*}

To achieve our objectives, we start by understanding the normal behavior of charging stations \textcolor{black}{by examining a real-life EVCSs dataset. This dataset was obtained from Hydro-Quebec as part of a legal agreement and research collaboration. Hydro-Quebec owns and operates, through a subsidiary, the public EVCSs in Quebec. This data is used to understand the behavior of the public EVCSs ensuring normal behavior and extract certain features that allow us to build our own realistic data-driven normal EVCS behavior.} First, we identify the state changes of a charging station. The charging station alternates between three states: idle, charging, and discharging. Whenever a charging station receives a charging request it transitions from idle to charging, and when it receives a stop charge request the charging station goes back to the idle state. However, the duration that the charging station spends in any of the states needs to be understood since the oscillatory load attack is tightly coupled with the total attack load and the time spent in each state. The dataset is acquired from 6,000 \textcolor{black}{EVCSs located in different geographical locations of Quebec from 2018 to early 2022 to cover all four seasons of Canada and their corresponding influence on charging behavior.} The data contains multiple principal metrics about charging sessions (e.g., start time, end time, and duration of charging). The normal behavior of the charging stations falls under two general observations 1) normal behavior of a charging station with charging > 5 minutes; 2) switching behavior of an individual charging station that does not impact the grid. \textcolor{black}{Consequently, we analyze the charging behavior of one heavily utilized and one lightly utilized EVCS located around the downtown area in Montreal}. The average duration of EVCS 1 (Figure \ref{fig:station1})is 24 minutes with a minimum of 55 seconds. Whereas EVCS 2 recorded an average duration of 8 minutes with a minimum of 26 seconds. After further analysis of EVCS 2 (Figure \ref{fig:station2}), a switching behavior is observed at 8:33 A.M which was followed by two other switches at 8:34 and 8:35. Similar behavior was repeated at 9:47, 15:21, 17:11, and 19:15. \textcolor{black}{The two charging station behavior patterns are identified based on their utilization where their hardware specifications are the same (providing an 11 kW charging rate)}. It is worth mentioning that, a switching behavior occurring simultaneously on numerous charging stations would be considered a coordinated oscillatory load attack. This observation shows that the behavior of a charging station by itself is not enough to detect oscillatory load attacks because it might cause numerous false positives and false negatives due to the presence of a switching behavior during the normal operation of a charging station. As the number of charging stations increase this phenomenon is expected to increase among charging stations. Moreover, since detection is occurring on the charging station that does not have any information about other charging stations, we couple the events happening on the charging station with the frequency readings over the studied t$_{1}$ time. The frequency is a global variable shared between all charging stations that are connected to the same bus, which allows the charging station to gain global knowledge of the EVCSs connected to the same bus while keeping the detection local to itself. During a synchronized oscillatory load attack, events that occur on the EVCSs are tightly coupled with grid behavior. Therefore, we couple the events that occur on the charging station (e.g., start time, end time, and duration) and the grid behavior (e.g., frequency) over time. Hence, using the mentioned features, we aim to detect and localize a synchronized oscillatory load attack with the fine granularity of identifying the charging stations that were compromised to perform such attacks.

\subsubsection{Data Synthesis and Collection}
\label{sec:datacollection}
A crucial part of our detection mechanism is creating a comprehensive and realistic dataset that resembles both normal and malicious behaviors. Since the existence of such attack data is scarce in real life and due to the unique features we chose, \textcolor{black}{we create a realistic data-driven EV load profile. To achieve this, we independently simulate a Poisson arrival process of EVs to each EVCS. The charging time of these EVs is then simulated as a truncated Gaussian distribution. The parameters of the arrival and charging time models are specified for different periods during the day and for different seasons. These parameters are tuned based on the Hydro-Quebec EVCS dataset. Finally, we also simulate the impact of normal charging on the power grid along with the behavior of the power grid as a result of the different oscillatory load attacks launched. The simulated dataset will be used to train our detection model due to the lack of real data with the required granularity (0.5 seconds) published online. We will focus on anomaly detection for the detection of synchronized oscillatory load attacks. To this end, we couple the behavior of the EVCSs and grid under normal and attack conditions}. As mentioned above, our method offers the coupling of the \textcolor{black}{cyber} events occurring on the charging station to the \textcolor{black}{physical data, i.e., power grid frequency behavior.}

The normal arrival of new charging requests at a charging station is coupled with the normal frequency behavior of the bus to which the charging station is connected. \textcolor{black}{The arrival of charging requests during an oscillatory load attack is coupled with the abnormal frequency behavior of the bus to which the charging station is connected. To this end}, the IEEE New England 39-bus system \cite{athay1979practical} was built in MATLAB Simulink to gather the required power grid data. \textcolor{black}{We use the MATLAB-Simulink 2020a Specialized Power Systems Toolbox which is widely used for system stability studies \cite{bhatt2017analysis}}. The Simulink Power System Toolbox allows us to model all the different components of the power system (i.e., loads, lines, transformers, generators, and generator control systems). Given the dynamic behavior of the power system, it is mostly governed by the control system of the generators and \textcolor{black}{we use the models commonly adopted by stability studies i.e.,} round rotor type synchronous machine block of Simulink, generator exciter model IEEE T1, turbine speed governor IEEE G2, and a power system stabilizer based on IEEE Std 421.5. \textcolor{black}{The simulations were performed with a simulation step size of 1ns}.

The implemented model would allow the study of the transient and steady-state behavior of the system. To simulate the normal frequency fluctuations of a power grid, we added random load blocks to all load buses. IEEE \textcolor{black}{grid models} have constant loads which usually represent the average load of the bus. However, real consumer behavior is random during a short span of a few minutes such that its average is what is reported and planned by utilities. This gives rise to the need to simulate small random perturbations in the loads of our power system which would lead to normal frequency variations. The random load blocks we added to all load buses are \textcolor{black}{constituted of a random number generator and a dynamic load block which are} provided in Simulink. The magnitude of the random number generator is scaled by the nominal load of the bus it is connected to and a multiplication block with a percentage cap that is changed in every simulation run. This setup is used to control the real and reactive power of the dynamic load block. The power factor of the random load block is maintained at 0.8 lagging to simulate benign consumer load variation. In half the simulations, the random source was set to follow a Gaussian distribution and in the other half, it followed a Uniform distribution to increase the randomness in our data and simulate close to real-life load perturbation. To avoid \textcolor{black}{the pattern effect of pseudo-random number generators and to insure true randomness}, we utilize the Mersenne Twister algorithm with a period length of $2^{19937}-1$ and we initiate the shuffle command before every simulation run to \textcolor{black}{randomnly select new seeds for the random number generator and guarantee further randomness}.

The constructed system is used to create a dataset of 5,000 normal (no attack) scenarios and 5,000 oscillatory load attack scenarios by collecting \textcolor{black}{cyber layer measurements, EVCS events, and physical layer measurements, grid frequency}, from the simulated system. Our normal attack dataset constitutes 80\% of a behavior similar to EVCS 1 in (Figure \ref{fig:station1}), whereas the other 20\% follow the behavior similar to EVCS 2 in (Figure \ref{fig:station2}). Moreover, we identify the following 4 scenarios and classify them as normal behavior:

\begin{itemize}[leftmargin=*]
   \item Very slow charging station switching (normal charging request start and stop) and normal bus frequency behavior. Charging events with a very low arrival rate (e.g., $\lambda < 6$ event per 60 minutes), while the grid shows a normal frequency fluctuation.
    \item Very slow charging station switching and abnormal bus frequency behavior. Charging events with a low arrival rate (e.g., $\lambda < 6$ event per 60 minutes), while the grid shows abnormal fluctuation in the frequency. This abnormal grid behavior can either result from some sudden benign disturbance on the grid or from an attack that does not involve the charging station in question.
    \item Slow charging station switching and normal bus frequency behavior. Charging events with a high arrival rate ($\lambda > 6$ events per 60 minutes), with a normal frequency fluctuation \textcolor{black}{as a result of normal consumer behavior}.
    \item Fast charging station switching and normal bus frequency behavior. Charging events with a very high arrival rate ($\lambda > 6$ per 60 seconds), coupled with normal frequency fluctuation. This case represents a few actual cases we monitored where the EV owner connects and disconnects a few times a minute at the time of arrival. Furthermore, the absence of any abnormal frequency behavior means the absence of any abnormal behavior on the power grid and thus the absence of a synchronized attack.
\end{itemize}

We identify two cases of attack behavior that include fast charging station switching with abnormal bus frequency where the adversary performs periodic attacks and record the events and the frequency fluctuation. The attack frequency was as fast as 1Hz \cite{sayed2021electric}. The second case is slow charging station switching with abnormal bus frequency where an adversarial attacker with more resources can compromise more charging stations than required for the attack and distribute the switching behavior among them to remain stealthier. If the attack for example requires $n$ charging stations switching at a frequency of $f$ Hz, the attacker can compromise $m \times n$ charging stations and switch them at a frequency of $\frac{f}{m}$ to obtain an identical aggregate effect but with a fraction of the switches on every charging station and remain stealthier.

The switching of the charging stations is crafted in a way such that the aggregate load on the buses \textcolor{black}{adheres to the following behavior}. For the slow switching attack scenario, we distribute the load such that the aggregate switching load has a duty cycle (the proportion of time during which an electrical device is operated) of 35\%, 50\%, and 60\% each constituting a third of all cases. Moreover, \textcolor{black}{all three oscillatory load attack variations are crafted} with an aggregate period between 1s and 2 s (0.5 -1 Hz) and an aggregate attack magnitude between 10\% and 30\% of the bus load. \textcolor{black}{This provides us with a comprehensive dataset that simulates the different types of oscillatory load attacks (switching and dynamic) that are discussed in Section \ref{sec:oscillatory}. The different types of attacks differ in their magnitude, periodicity, time of the attack, and stealthiness; however, their collective impact on the power grid is comparable in terms of forcing abnormal frequency oscillations. Consequently, in this work, we aim at detecting oscillatory load attacks to mitigate their impact on the power grid}.

Given the obtained dataset of a \textcolor{black}{cyber events on the charging station and with frequency data} from the grid, we adopt a time-series selection/representation approach. After coupling the two features that vary with time, where one of these features (events on charging station) affects the other feature (frequency) it transforms our problem from a time series classification to a multivariate time-series classification with two axes of difficulty. Temporal and spatial relationships are learned and mined using deep learning techniques to extract the variation of the features with respect to time and how features vary with respect to each other.

Each charging station monitors the previous 120 seconds in \textcolor{black}{rolling windows} whenever it receives a request and fetches readings from its log file. The events and frequency are recorded every 0.5 seconds, which results in 240 readings for each feature per instance. We fix the \textcolor{black}{rolling window} size and the number of readings, due to simulation environment limitations of 120, and 240 respectively. In our study, we devise deep learning models to detect attacks and\textcolor{black}{optimize them to detect attacks after} 5 seconds and 10 seconds \textcolor{black}{of attack start}, which we call 5-Attack and 10-Attack, respectively. This means that the last 5 seconds or 10 seconds of the 120-second window will have attack features whereas the rest would be normal behavior. \textcolor{black}{Through this approach, our detection mechanism can successfully detect almost all attacks as early as 5 seconds by only viewing the beginning of the attack. Since our methodology is based on a rolling window, this allows the algorithm to identify attacks if any false negatives occur in previous windows. Furthermore, since we are utilizing a rolling window, our algorithm can start detecting the attacks, with some degree of success, as soon as 1s after the attack starts. Decreasing the size of the attack windows we are optimizing for would decrease the accuracy of the machine/deep learning model as stealthy attack scenarios would be misclassified. In the stealthy version of the attacks, the attacker launches slow oscillatory attacks as well as distributes the switching among multiple charging stations. This means that in short windows of time, the charging station behavior would look normal since only one event or possibly no events at all occur making their behavior look completely normal. Consequently, we chose to optimize for 5-second attack windows to detect a wide variety of attacks without compromising on accuracy. Our window rolls in intervals of 1s (split into two 0.5s sub-intervals) which means that our detector can start recognizing the attacks within 1s of their attack start.}

\subsubsection{Feature Selection}
Given the obtained dataset of events coupled with their power grid frequency readings, we adopt two feature selection/representation approaches. We collect a sequence of observations that are taken sequentially in time, which defines our data to be time series. Consequently, to use a set of time series $\mathscr{D} = {\mathscr{X}_{i=1}^{N}}$ as input for the deep learning algorithms, we map each time series $\mathscr{X_{i}}$ of set $\mathscr{D}$ into a matrix of $\mathscr{N}$ rows and $\mathscr{M}$ columns by choosing $\mathscr{M}$ data points of the two variables (events and frequency) from each time series $\mathscr{X}_{i}$ as elements of the feature vector. This allows the deep learning model to take into account temporal and spatial information and find the correlation between the events and the frequency.

\subsubsection{Classification Models}
\textcolor{black}{Given the features we selected and the complexity of relating cyber data and physical data to perform the classification, we need to implement a classification algorithm able to handle this data and preserve its properties}.To this end, we implement and evaluate different Deep Learning classification models. Specifically, we use Recurrent Neural Networks (RNN) to capture the order, occurrences, and structure of the events. We leverage a special type of RNN, namely Long Short Term Memory (LSTM). LSTMs preserve the errors that will be backpropagated through layers that allow LSTMs to continue learning over many time steps. LSTM is unique in its capability to learn what information to store in long-term memory. LSTM also allows the neural network to identify patterns and sequences in the data by learning temporal relationships between multiple time steps while utilizing memory gates. \textcolor{black}{These features allow LSTM to capture the temporal relations between events in a multivariate time-series data classification problem}.

We further explore a special type of Convolutional Neural Network, namely a 2D Convolutional LSTM. In our multivariate time series classification, \textcolor{black}{it is important to capture spatial interpretation and relationships}. The events that occur on the charging station, in case of synchronized switching attacks, are tightly coupled with the power grid behavior. \textcolor{black}{Capturing spatial information between the cyber layer and physical layer features allows the algorithm to capture the correlation between events on the EVCS and the power grid frequency behavior}. Thus, ConvLSTM nodes possess convolutional capabilities to handle spatial information and LSTM capabilities to handle temporal information, solving our dual-axis data relationships \cite{shi2015convolutional,zhang2022review}. We use ConvLSTMs to overcome the major limitation of LSTMs in finding spatial relationships between features over multiple time steps \cite{shi2015convolutional}. \textcolor{black}{Unlike LSTM which flattens the data and loses any spatial relationships}, ConvLSTM replaces \textcolor{black}{the LSTM gate in each LSTM cell with convolution operation made up of several filters of square matrix kernels}. By doing so, \textcolor{black}{ConvLSTM} captures underlying spatial features by convolution operations in multiple-dimensional data \textcolor{black}{while preserving the temporal relationship between the data as well}. 

\subsubsection{Model Evaluation and Comparison}
We follow several standard methods to evaluate the overall effectiveness of the implemented classification models to compare their outcomes. More specifically we use metrics such as accuracy, recall, precision, and F-measure. Moreover, we use the confusion matrix, which is a useful method for discussing the effectiveness of the implemented deep learning models. The confusion matrix shows the number of data instances that were classified correctly using the model (true positive and true negative) and the number of data instances that were misclassified by the model (false negative, false positive).

\subsection{\textcolor{black}{Distributed Mitigation Methodology}}
\label{sec:mitigation}
In this paper, we also aim to mitigate the impact of oscillatory load attacks in a distributed \textcolor{black}{and lightweight} manner and assist the power grid to \textcolor{black}{return to its normal state} easily.

\begin{algorithm}
\caption{\textcolor{black}{Algorithm describing the conceptual model of the detection and mitigation mechanisms}}
\label{alg:two}

\SetKwInOut{Input}{Inputs}
\SetKwInOut{Output}{Output}
\SetKwBlock{DoParallel}{do in parallel}{end}
\SetKwComment{Comment}{//}{ }

\Input{$CS_{log}$:  Charging station logs,\Comment*[r]{Events and frequency logs}\\
$M_{1}$: the 5-Attack trained deep learning model,\Comment*[r]{model to detect attacks within the first 5 seconds}\\ 
$M_{2}$: the 10-Attack deep learning model \Comment*[r]{model to detect attacks within the first 10 seconds}}

\Output{$L_{test}$: the prediction class for the test sample in $CS_{log}$.\\
$d_{1}$: the delay of the incoming requests.}
\While{True}{
    \ForEach {Event $e_{i} (t) \in CS_{log}$}{
        \DoParallel{
        $L_{1} \gets $ Prediciton($M_{1}$, $x_{i}$)\Comment*[r]{predict the class of the behavior recorded}

            \While{L$_{1}$ or L$_{2}$ is Abnormal}{
                $d_{1} \gets $ Random Delay$_{0 < d \leq 4 seconds}$ \Comment*[r]{continue generating random delays to the new incoming requests}
                Report abnormal behavior to operator/utility
            }
        }
    
        \DoParallel{
        $L_{2} \gets Prediciton(M_{2}, x_{i})$ \Comment*[r]{ apply the 10-attack deep learning model to decrease false negatives}

            \While{L$_{1}$ or $L_{2}$ are Abnormal}{
                $d_{1} \gets $ Random Delay$_{0 < d \leq 4 seconds}$\Comment*[r]{If any of them is abnormal add a delay}
                Report abnormal behavior to operator/utility
            }
        }
        \If {L$_{1}$ and L$_{2}$ is not Abnormal}{
            $d_{1} \gets $ 0\Comment*[r]{If the prediction of both models did not show abnormal behavior then stop the mitigation mechanism}
        }
    }
}
\end{algorithm}
Consequently, in this section, we discuss and evaluate a lightweight and distributed mitigation mechanism against oscillatory load attacks. After locally detecting the attacks within 5 seconds on an EVCS as discussed in the previous sections, a charging station can either discard a request or create a random delay by taking this decision independently. \textcolor{black}{In \cite{kabir2021two}, Kabir et al. proposed a centralized physical-layer mitigation technique that requires the utility to upgrade its existing generators with new control mechanisms. While the oscillations were successfully damped using their method, the oscillations on the power grid were never completely eliminated}. Moreover, a centralized cyber-layer mitigation technique can only mitigate attacks launched by public charging stations. Due to the aforementioned limitations, we propose a lightweight and distributed mitigation mechanism which can be deployed on the charging station itself (public and/or private). After detecting attacks, each station independently creates a random delay for all incoming requests to break the attacker's synchronization \textcolor{black}{ability} and hence minimize the impact on the grid after a \textcolor{black}{persistent} attack. \textcolor{black}{Our main goal is to create a lightweight and distributed cyber-layer mitigation mechanism that aligns with the EVCS ecosystem deployment. The charging stations are characterized by low computing power which motivated the need for an independent mitigation mechanism. Consequently, we studied the effectiveness of using the random delay to mitigate the impact of forced oscillations on the power grid. In future work, with the aim to minimize the delay, we plan to utilize deep learning techniques that tailor the delay to each attack type based on the behavior and the load on the grid.} \textcolor{black}{We describe in Algorithm \ref{alg:two}, the conceptual model of the algorithm and how it is integrated with the mitigation mechanism. The detection is an online algorithm that is running in real-time to detect oscillatory load attacks. We run the 5-attack detection model to detect attacks early and the 10-attack detection model to detect any false negative data samples that resulted from the first step. We utilize this two-step continuous detection technique to lower the false negative impact on the power grid and provide continuous monitoring over the ecosystem, when the first detection technique detects an attack it creates a random delay for any new request ranging between 0 and 4 seconds. However, this delay is removed if the 5-attack and 10-attack deep learning models stop classifying the rolling windows as attacks. Consequently, we can then utilize the 10-attack model to minimize the number of false negatives that might have been misclassified by the 5-attack deep learning model}.

\begin{figure}[t]
    \centering
    \includegraphics[width=0.9\linewidth]{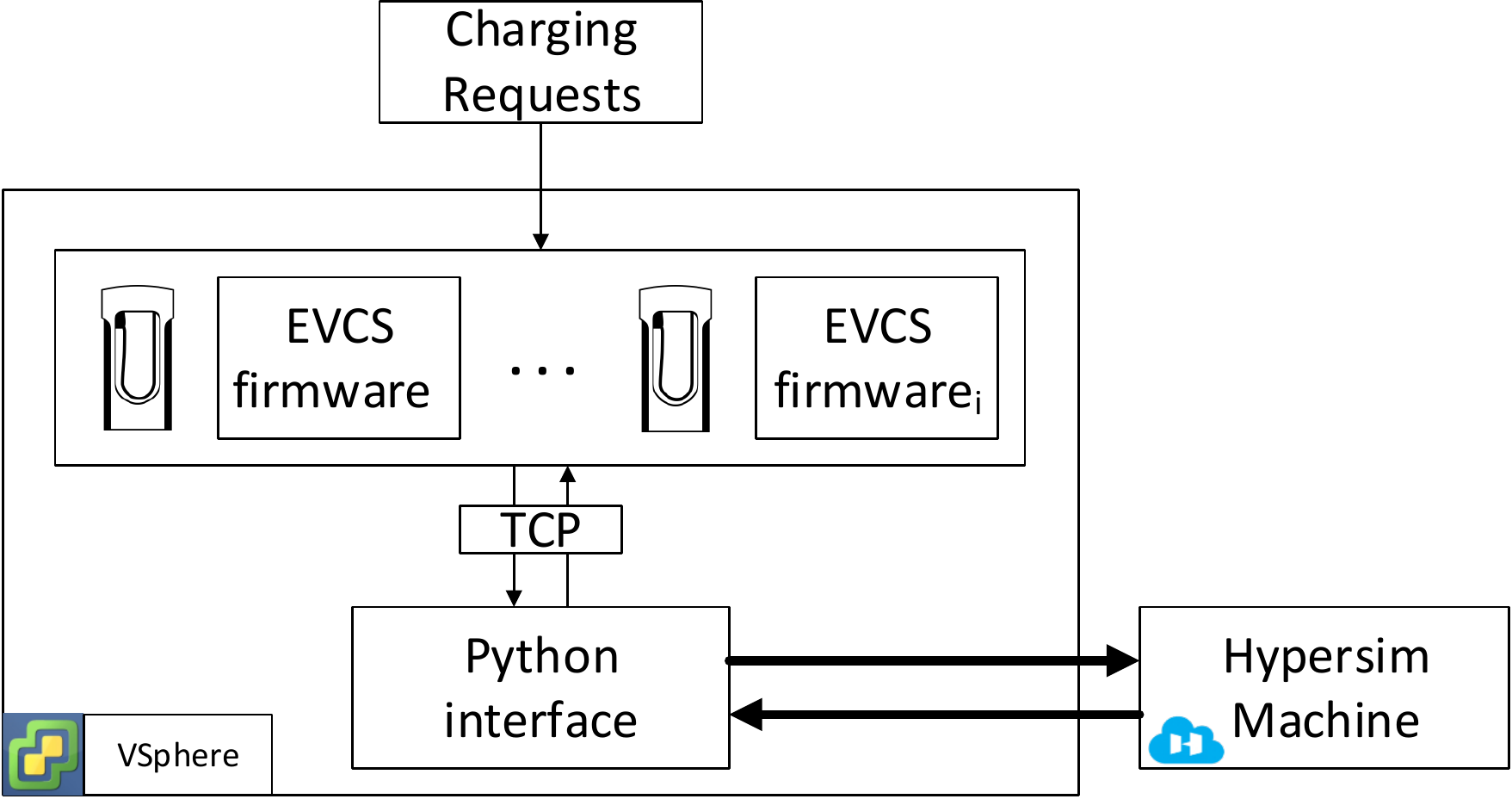}
    \caption{Co-simulation architecture}
    \label{fig:co-sim}
\end{figure}

\subsubsection{Test-bed}

As part of our efforts to study the EV ecosystem, we create an EV co-simulation platform that integrates the different components to simulate the cyber and physical layers of this ecosystem. The cyber layer is composed of the mobile application, central management system \textcolor{black}{(OCPP server)}, and the charging station cyber interface \textcolor{black}{(OCPP client and firmware). The} physical layer is composed of the charging station's physical interface and the power grid.

We simulate the cyber-layer using vSphere \cite{marshall2018mastering}, which is a VMware virtualization platform. vSphere is built from VMware ESXi, a Type 1 hypervisor that is responsible for abstracting processors, memory, storage, and other resources into multiple virtual machines, and a VMware vCenter Server, that allows us to manage and control virtual machines. Moreover, we utilize Hypersim, which is a real-time power system simulator, that helps model power grids and run real-time simulations on dedicated \textcolor{black}{multi-processor hardware (OpalTagert) and is connected to the Hypersim VM over a Local Area Network}.

\begin{table}
\caption{ \textcolor{black}{Specifications of the real-time co-simulation testbed.}}
\begin{center}
\begin{tabular}{ c|c } 
 \hline
 Technology & Specification \\ \hline\hline
 vSphere ESXi & Version 6.0.0\\\hline
 Hypersim & Version 2022.1\\\hline
 Hypersim Simulation Step Size & 25\textmu s\\\hline
 OpalTarget & OP5707XG - RCP/HIL Virtex-7 FPGA-based Real-Time Simulator\\\hline
 EVCS VM &  1GB 1 CPU\\ \hline
 CMS VM & 1GB 1 CPU\\\hline
 Mobile app VM & 1GB 1 CPU\\\hline
 Python Interface & Python 3.7\\\hline
 
\end{tabular}
\label{table:co_sim_parameters}
\end{center}

\end{table}
The architecture of our testbed is illustrated in Figure \ref{fig:co-sim} \textcolor{black}{and the specifications of the testbed elements are stated in Table \ref{table:co_sim_parameters}}. We simulate requests (start and stop) sent to multiple charging stations. The charging stations during operation are \textcolor{black}{modeled as the dynamic load blocks in Hypersim connected to their respective buses in our} power grid. Consequently, the requests sent to charging stations are then aggregated in real time based on the bus that they are connected to. Then we create a load profile that resembles the changes over time. The aggregated load profile is read by the python interface \textcolor{black}{in real-time}. The python interface establishes a session with Hypersim then controls and executes the load profile to simulate the different loads on various buses.

\section{Experimental Results}
\label{sec:results}
As described in Section \ref{sec:datacollection}, we focus on oscillatory load attacks initiated by exploiting vulnerabilities in the EV charging ecosystem. We study periodic attacks and stealthy attacks where an attacker with enough resources can group charging stations and alternate the switching between different groups that leads to inconspicuous behavior on the charging station.

\subsection{Distributed Detection Mechanism Results}
We use our multivariate-time series representation by taking the temporal and spatial relationship of the two features (events and frequency) into consideration. We implement several deep learning models such as LSTM for temporal relationships and ConvLSTM from Spatio-temporal relationships between the multivariate time series. We test our models on 5-Attacks and 10-Attacks.

\textsc{Data Pre-Processing}:
To feed our data to the deep learning algorithms, we encoded our features by converting the events (start and stop) to numerical values. Moreover, we normalize the frequency readings by re-scaling the data and fitting all the frequency data points between 0 and 1. To preserve the shape of the original distribution and the information embedded in it, we can represent the normalization as follows:

\begin{equation}
    x_{i normalized} = \dfrac{x_{i} - min(X)}{max(X) - min(X)}
\end{equation} 

Where $x_{i}$ is any value from the feature $x$ (e.g., frequency), $min(X)$ is the minimum value from the feature, and $max(X)$ is the maximum value of the feature. We utilized MinMaxScaler to normalize our frequency feature vectors and obtain x$_{i normalized}$. Finally, we map each class (normal and abnormal), using binary label encoding, to 0 or 1. After that, we split our data into training (80\%) and testing (20\%) \textcolor{black}{subsets}.

\textsc{Model Selection and Evaluation}:
We studied different deep learning models to classify oscillatory load attacks based on behavioral events on the charging station and their consequent effect on the power grid as represented by the frequency readings. The structure and layers of these models are described in the following sub-sections, along with the evaluation results. Moreover, we chose the Adam Optimizer\cite{adamoptimizer} for this classification problem. Adam outperforms other optimizers, such as Root Mean Square Propagation (RMSprop) and Adaptive Gradient Algorithm (AdaGrad), because of its bias-correction which helps Adam towards the end of the optimization as gradients become sparser \cite{ruder2016overview}. We systematically enhance our outcomes, by iteratively adding layers to a simple model (fewer layers) until we reached a relatively good fit (no under-fitting or over-fitting). Consequently, we perform hyper-parameter tuning using the Random Search algorithm. Finally, we identified the hyperparameters that yielded the highest F-measure among the runs. The parameters tuned are the learning rate in the Adam optimizer, the proportion of drops, the number of neurons in a layer, the size of batches, the number of epochs, filter size, and kernel size. Finally, we evaluate the speed of each model as a measure of their computational performance, and the training time to measure the complexity of the model.

\textsc{Hyper-Parameter Tuning and Applied Random Search}:
It is worth mentioning different techniques were applied to decrease overfitting and achieve a good fit on training data. We used dropping techniques, which refer to dropping out/ignoring units (i.e., neurons) during the training phase with a certain probability \cite{srivastava2014dropout}. Moreover, we use the batch normalization technique to stabilize the distribution of inputs (over a mini-batch) to a given layer during training. This helps in dramatically reducing the number of training epochs required to train deep networks \cite{santurkar2018does}. We applied hyper-parameter tuning to find the best parameters to achieve a good fit since there exists a large number of variables that can be tuned to enhance the training. Finally, we use binary cross-entropy as a loss function. 

For each model, we applied a Random Search and a refined Random Search algorithms \textcolor{black}{to tune the hyperparameters that will yield the best results}. In a deep learning model, various parameters contribute to finding the best fit for the training data (e.g., learning rate, decay, batch size, etc.). \textcolor{black}{The Random Search algorithm uses a random combination of these values and trains the deep learning model on all these combinations. In the first step of our 2-step Random Search, we generated 500 different combinations of these Hyperparameters and selected the combination that resulted in the highest F1 score. The random search uses all of these 500 combinations and trains the deep learning model 500 times. We then select the model and the parameters that achieved the highest F1 score. For refined results, we apply the second step of (refined) Random Search by generating 100 different combinations of hyperparameters in the 10\% realm of the ones found in the first step, to try enhancing our results further.} The random search performs similarly to meta-heuristics and grid search, however, with a lower computational cost \cite{mantovani2015effectiveness}. 

In the following subsections, we present the analysis of the performance of the deep learning algorithms that were implemented to detect switching attacks. We select the deep learning model which shows the highest F-measure score, as it is more representative of the false negatives and false positives in the data. Moreover, we also base our judgment on the number of false-negative, where an attack is misclassified as normal.
        


\begin{figure}[t]
    \centering
    \includegraphics[width=0.9\linewidth]{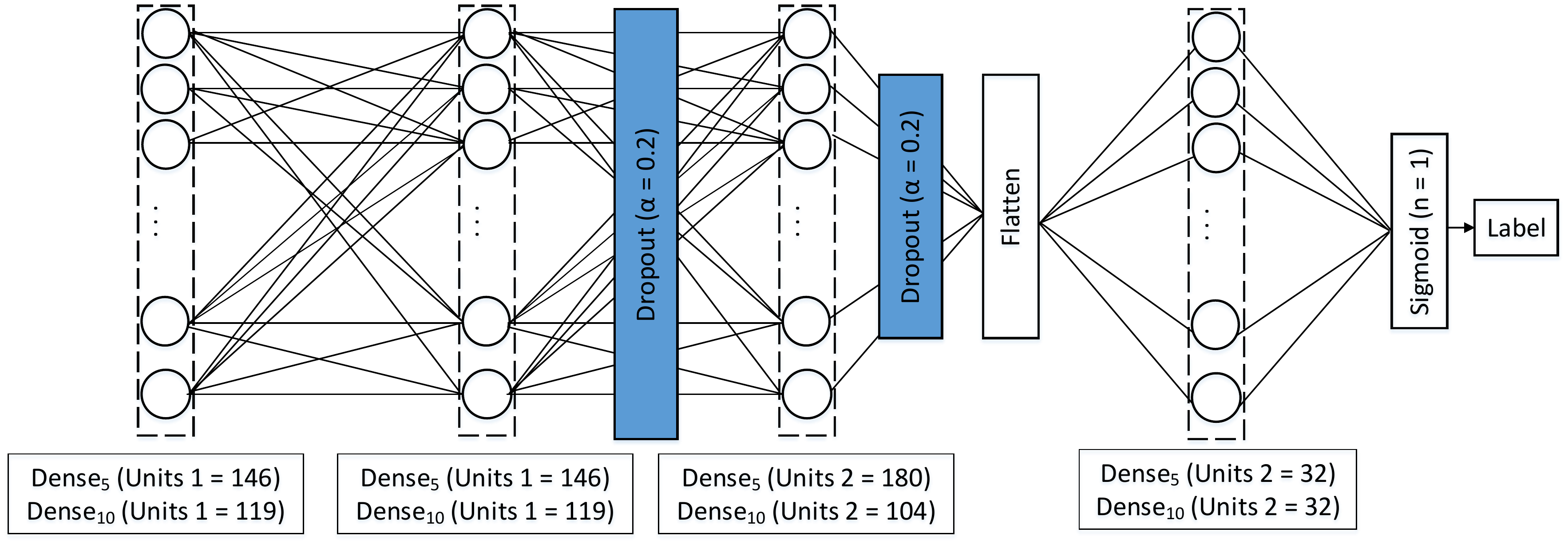}
    \caption{Structure of the Long-Short Term Memory Model. } 
    \label{fig:lstm}
\end{figure}

\textsc{Long-Short Term Memory (LSTM)}: We use LSTM model as a benchmark against other spatiotemporal deep learning models. The developed LSTM architecture is depicted in Figure \ref{fig:lstm} and consists of the following layers:

\begin{itemize}[leftmargin=*]
    \item Input Layer: The input of the network is 240 x 2 of encoded events and frequency reading collected over time.
    
    \item LSTM Layer: This is the main building block of an LSTM deep neural network and is responsible for learning the order dependency in our feature space.
    
    \item Fully Connected Layers 1 and 2: After the LSTM layer, we add a fully connected layer with a Leaky ReLU activation function. To decrease overfitting we apply batch normalization along with a dropout. The output is then fed into another fully connected layer with a Leaky Relu activation function. The Leaky ReLU activation function was developed to overcome one of the major shortcomings of the ReLU activation function. The ReLU activation function faces the "Dead ReLU'' issue that occurs during backpropagation when no learning happens as the new weight remains equal to the old weight. Followed by batch normalization and a dropout layer. The output of these layers is three-dimensional, consequently, a copy of the output is collapsed into one dimension.
    
    \item Fully Connected Layer 3: The one-dimensional output is then fed to a connected layer initialized by a truncated normal distribution. Consequently, the last layer combines the features learned in previous layers and applies a Sigmoid function to output a probability between 0 and 1 that indicated the class of the data samples.
\end{itemize}

After running our random search (500 runs), we achieve a relatively good accuracy (97.5\%) and F-measure scores (97.493\%) on the 5-Attack dataset. Moreover, we run a 500 runs random search to tune the LSTM model on the 10-Attack dataset which achieved a better accuracy (99.4\%) and F-measure scores (99.405\%) as shown in Table \ref{table:outcomes}. In our experiments, it is crucial to look at how well our model classified attacks. The confusion matrix (Table \ref{table:confusionMatrix}) shows that using the 5-Attack dataset, 972 attack samples were classified correctly, whereas 40 data samples were misclassified. Moreover, using the 10-Attack dataset, 1002 attacks out of the 1012 attacks were classified correctly using our novel Long Short-Term Memory deep learning model, and around 1\% of the attacks were incorrectly classified. This confirms that oscillatory load attacks need more in-depth analysis to improve the accuracy of the model especially since the 5-Attack dataset achieved a high false negative (3\%). It is worth mentioning, the smaller the attack window viewed by the detection system the earlier we detect attacks. Moreover, we note that the behavior of charging stations in normal conditions has some similarities with charging stations under stealthy attacks. The attacker, in stealthy attacks, tries to mimic the normal behavior of a charging station by dividing the switching behavior across different groups of charging stations and alternating the switching between them, which could be the reason for such misclassification. Although the misclassification is not huge, the impact that might occur out of successful attacks would cause a devastating impact on the power grid that could lead to tripping lines and overloading generators.

We evaluate our models based on the training time and the time to make a prediction on the 20\% test set provided. The training time of the best-performing LSTM deep learning model is ~4 minutes for both datasets. The time to train the model is a good indicator of the complexity of the model and the resources needed for future enhancements. Moreover, the time to predict the labels of our 2000 sample test set for the 5-Attack dataset and 10-Attack dataset is 6 and 12 seconds, respectively, which equates to 0.003 seconds and 0.006 seconds on average for each data sample. The speed of each model is a measure of its computational performance.

\begin{figure}[t]
    \centering
    \includegraphics[width=0.9\linewidth]{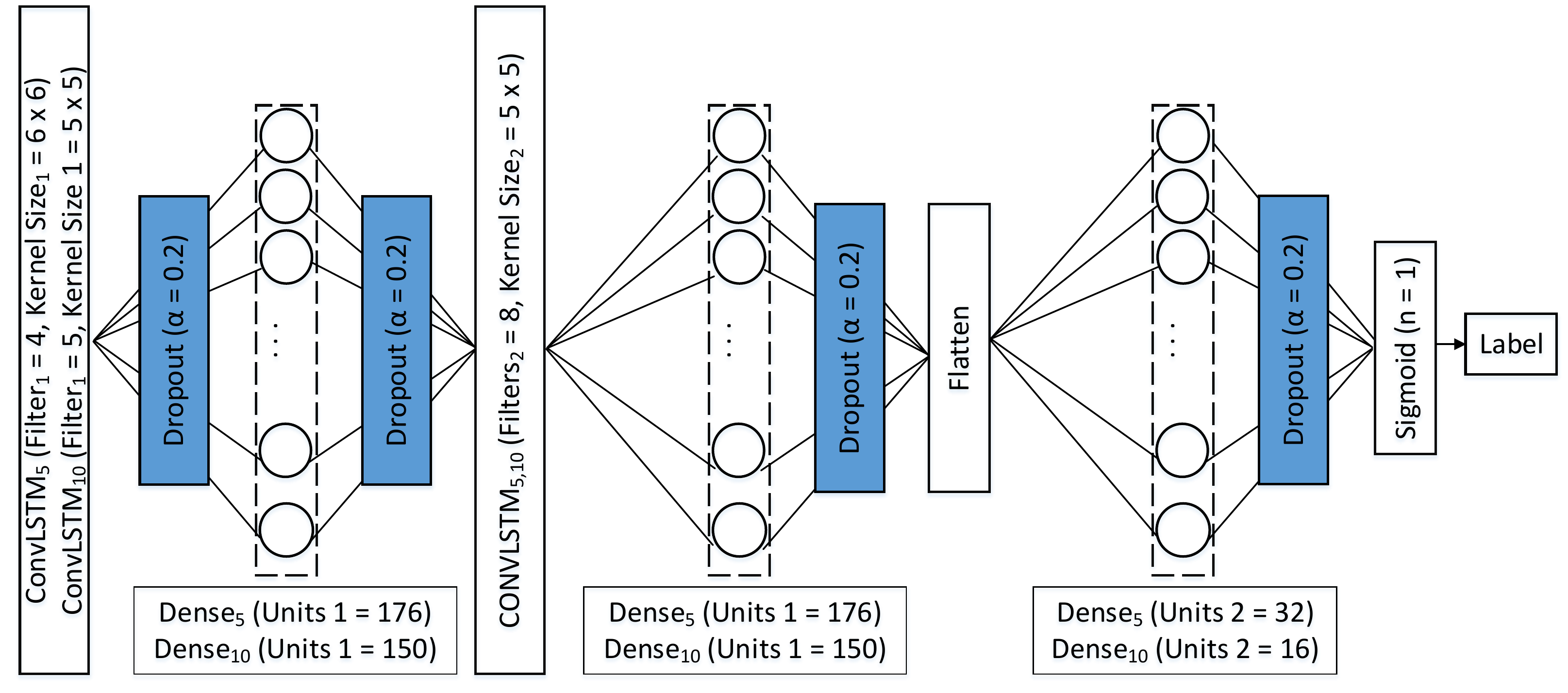}
    \caption{Structure of the Convolutional Long-Short Term Memory Model.}
    \label{fig:ConvLSTM}
\end{figure}

\begin{table}[t]
    \centering
    \caption{\textsc{Optimized Hyper-Parameters for the Implemented Models}}
    \begin{tabular}{c | c c | c c}\toprule
        \multicolumn{1}{c}{} & \multicolumn{2}{c}{5-Attack} & \multicolumn{2}{c}{10-Attack}\\\midrule
        \multicolumn{1}{c}{} & \multicolumn{1}{c}{LSTM} & \multicolumn{1}{c}{ConvLSTM$_{2D}$} & \multicolumn{1}{c}{LSTM} & \multicolumn{1}{c}{ConvLSTM$_{2D}$}\\\hline
        \textbf{Learning Rate} & 0.014717& 0.0001939&0.00070810 &0.0001\\
        \textbf{Drops} & 0.34 & 0.2 & 0.2 & 0.18\\
        \textbf{Batches} & 56 & 30 & 40 & 34\\
        \textbf{Units 1} & 146 & 150 & 119 & 176\\
        \textbf{Units 2} & 180 & 32 & 104 & 16\\
        \textbf{Units 3} & 32 & - & 32 & -\\
        \textbf{Epochs}  & 5 & 6 & 6 & 7\\
        \textbf{Filter 1}& - & 4 & - & 5\\
        \textbf{Kernel Size 1}& -& 6x6 & - & 5x5\\
        \textbf{Filter 2} & - & 8 & - & 8\\
         \textbf{Kernel Size 2}& - & 5x5 & - & 5x5 \\\bottomrule
    \end{tabular}
    \label{table:Hyperparameters}
\end{table}

\textsc{Convolutional Long-Short Term Memory (ConvLSTM2D)}: We use ConvLSTM which is an LSTM variant. ConvLSTM is a type of recurrent neural network for multivariate spatiotemporal detection. It has convolutional structures that combine the ability of a convolutional neural network to incorporate spatial and temporal correlations into modeling and automatically capture the shared structures across variables (events and frequency). The developed ConvLSTM architecture, depicted in Figure \ref{fig:ConvLSTM}, consists of the following layers:

\begin{table}[t]
    \centering
    \caption{\textsc{Confusion Matrices for LSTM and ConvLSTM}}
        \begin{tabular}{c | c c | c c | c c | c c }\toprule
            \multicolumn{1}{c}{} & \multicolumn{4}{c}{5-Attack} & \multicolumn{4}{c}{10-Attack}\\\midrule
            
            \multicolumn{1}{c}{} & \multicolumn{2}{c}{LSTM} & \multicolumn{2}{c}{ConvLSTM$_{2D}$} & \multicolumn{2}{c}{LSTM} & \multicolumn{2}{c}{ConvLSTM$_{2D}$}\\\hline
             & N & A & N & A & N & A & N & A \\
            N & 978 & 10 & 985 & 3 & 986 & 2 & 985 & 3\\
            A & 40 & 972 & 9 & 1003 & 10 & 1002 & 1 & 1011\\\bottomrule
        \end{tabular}
    \label{table:confusionMatrix}
\end{table}

\begin{table}[t]
\centering
    \caption{\textsc{Classifiers Outcomes}}
    \begin{tabular}{c | c c | c c }\toprule
        \multicolumn{1}{c}{} & \multicolumn{2}{c}{5-Attack} & \multicolumn{2}{c}{10-Attack}\\\midrule
        
         \multicolumn{1}{c}{}& LSTM & ConvLSTM$_{2D}$ & LSTM & ConvLSTM$_{2D}$\\\hline
            Accuracy & 97.500 & 99.400 & 99.400 & 99.800  \\
            F-measure & 97.493 & 99.405 & 99.405 & 99.803 \\
            Recall & 96.047 & 99.111 & 99.012 & 99.901 \\
            Precision & 98.982 & 99.702 & 99.801 & 99.704 \\\bottomrule
    \end{tabular}
    \label{table:outcomes}
\end{table}

\begin{table}[t]
    \caption{\textsc{Classifiers Time}}
    \begin{center}

        
           \begin{tabular}{c | c c | c c }\toprule
        \multicolumn{1}{c}{} & \multicolumn{2}{c}{5-Attack} & \multicolumn{2}{c}{10-Attack}\\\midrule
        
         \multicolumn{1}{c}{}& LSTM & ConvLSTM$_{2D}$ & LSTM & ConvLSTM$_{2D}$\\\hline
        Training Time & 0:04:25 & 1:45:28 & 00:04:31 & 02:27:53   \\
        Prediction Time (s) & 0.003 & 0.011 & 0.006 & 0.014 \\\bottomrule
        \end{tabular}
    \end{center}
    \label{table:time}
\end{table}

\begin{itemize}[leftmargin=*]
    \item Input Layer: The input of the network is 240 x 2 of encoded events and frequency reading collected over time.
    \item ConvLSTM Layer 1: This is the main building block of the ConvLSTM deep learning network and is responsible for finding Spatio-temporal relationships in the multivariate time series. To decrease overfitting we apply batch normalization and a dropout.
    \item Fully Connected Layer 1: After the ConvLSTM layer, the output is fed into a fully-connected layer with a leaky ReLU activation function and followed by batch normalization and dropout.
    \item ConvLSTM Layer 2: The output is then fed into a second convolutional LSTM layer to derive further Spatiotemporal correlations from the features. Consequently, followed by batch normalization and a dropout layer. Moreover, the output is then flattened into a one-dimensional vector of numbers. 
    \item Fully Connected Layer 2: The one-dimensional output after flattening is then inputted into a fully connected layer initialized by a truncated normal distribution. Consequently, the last layer combines the learned weights in previous layers and applies a Sigmoid function to output a probability between 0 and 1 that indicates the class of the data samples.
\end{itemize}

After running our random search (500 runs) on the 5-Attack dataset, we achieve good accuracy (99.4\%) and F-measure scores (99.393\%). Moreover, we also test our deep learning model on the 10-Attack dataset and achieved better accuracy (99.8\%) and F-measure score (99.803\%). The confusion matrix for both datasets of the ConvLSTM model is depicted in Table \ref{table:confusionMatrix}. The classifier was able to correctly classify 985 out of 988 normal samples and 1002 out of 1012 attack samples. The performance of the classifier improved as we increase the attack window in the 10-Attack dataset. This shows that 0.99\% and 0.099\% of the attacks were misclassified for both datasets, as compared to the LSTM that achieved a 3.952\% and 0.99\% false negative rate. It is important to note that the number of misclassified attacks (false negatives) is an important indicator in choosing the best model to detect oscillatory load attacks. The impact of oscillatory load attacks initiated by the EVCS ecosystem has been studied by Sayed et al. \cite{sayed2021electric}. The risk of oscillatory load attacks is increasing as a result of the rapid deployment of charging stations. We acknowledge that the current deployment (number) of charging stations does not allow attackers to impact the power grid, however, with the current advancement and push towards electrifying the transportation system that is being enforced by governments such attacks will entail great risk.

Consequently, we evaluate the training time of the best ConvLSTM model accumulated to 2 hours approximately. The time consumed during training is substantial compared to the LSTM. This result is expected due to the increase in the number of training parameters tuned (e.g., filter and kernel sizes) that will allow the model to perform convolutional techniques on the input data to extract spatiotemporal relationships. The LSTM variant is labor-intensive in terms of training. However, the computational time of our ConvLSTM model is 22 and 28 seconds which amounts to 0.011 and 0.014 seconds on average per data sample for the 5-Attack and 10-Attack datasets.

\subsection{\textcolor{black}{Distributed Mitigation Results}}

To evaluate our distributed mitigation mechanism, we launch various oscillatory load attacks and study the impact on the grid on the 9-Bus system, which is \textcolor{black}{a simplified abstraction of the Western System Coordinating Council (WSCC) \cite{illinois} grid in North America}. In our test-bed setup, we are restrained to the 9-Bus system however, the general behavior of the different power grids is similar. Thus, our mitigation mechanism is easily reproducible on different power grids. After detecting the attacks within 5 seconds, every charging station adds a random delay to every request with the aim to deprive the adversary of the ability to synchronize attacks on multiple charging stations. Random delays are introduced with a maximum of 4 seconds. We chose 4 seconds based on the sensitivity analysis performed by Galletta et al. \cite{galletta2004web}, which implies a decrease in user performance and thus behavioral intentions begin to flatten when the delays extend to 4 seconds or longer in web interfaces. Kabir et al. \cite{kabir2021two} suggested discarding a request if an attack is detected. However, if a false positive attack was detected, the quality of service with respect to a valid customer is affected, which would lead to user frustration. As a consequence, we use a random delay topped at 4 seconds to preserve the quality of service.

\begin{figure}
     \centering
     \begin{subfigure}[b]{0.49\textwidth}
         \centering
         \includegraphics[width=\textwidth]{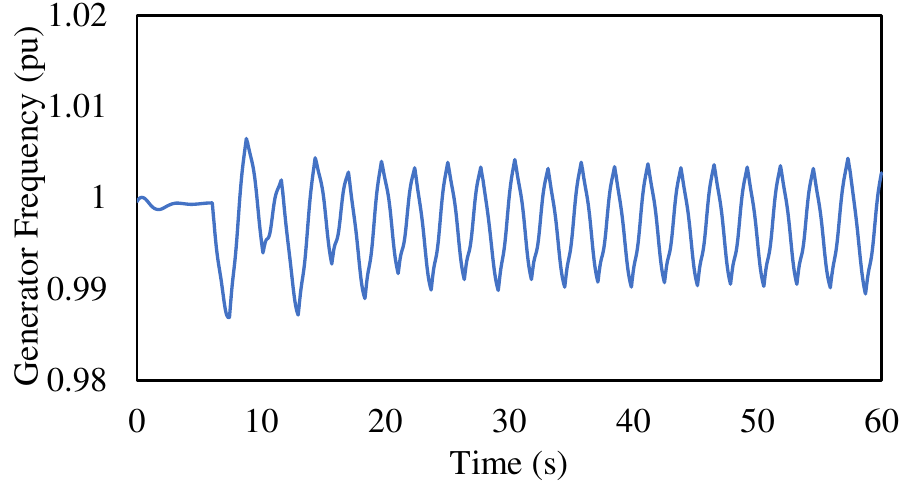}
         \caption{Generator 1}
     \end{subfigure}
     \hfill
     \begin{subfigure}[b]{0.49\textwidth}
         \centering
         \includegraphics[width=\textwidth]{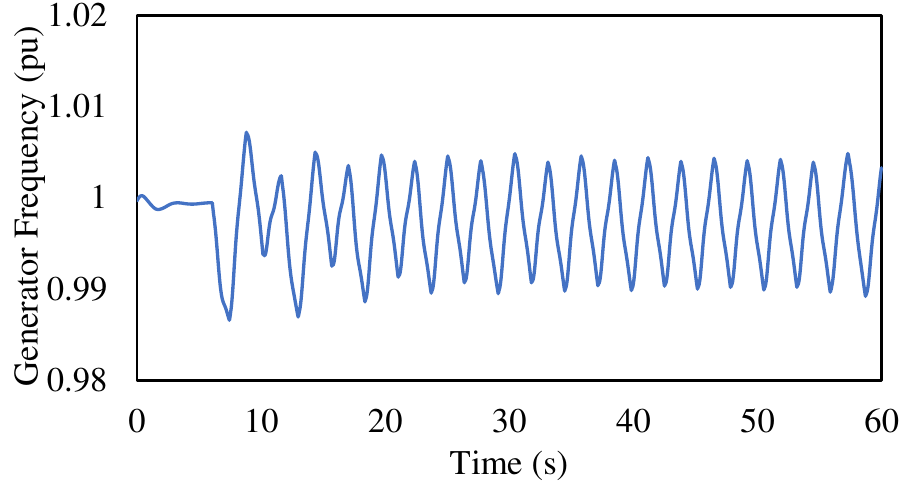}
         \caption{Generator 2}
     \end{subfigure}
     \hfill
     \begin{subfigure}[b]{0.49\textwidth}
         \centering
         \includegraphics[width=\textwidth]{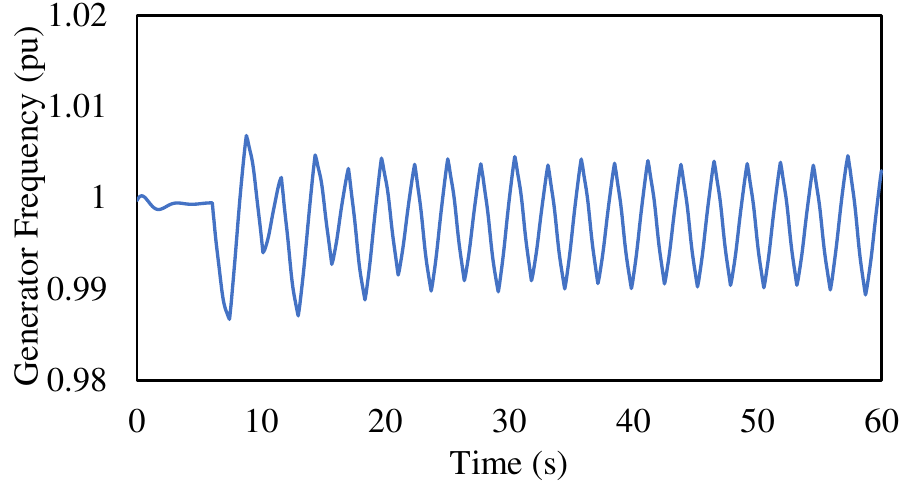}
         \caption{Generator 3}
     \end{subfigure}
        \caption{The variation of the generator's speed as a result of an oscillatory load attack.}
        \label{fig:speedAtt}
\end{figure}

In what follows, we demonstrate \textcolor{black}{an EV attack equivalent to 84MW load on one bus. This attack} is equivalent to about 7636 EVs charging at the 11kW Level 2 chargers. \textcolor{black}{Although such a number might be relatively high}, the growth in the EV numbers will soon provide a large enough surface to make it possible \cite{sayed2021electric}. \textcolor{black}{Relative to today's average charging rate of 24kW, the attacker would only need to compromise 3500 EVCS}. Moreover, as the EVCS market moves towards wide adoption of level 3 chargers, the number of needed compromised charging stations decreases. Level 3 chargers are DC fast chargers that deliver a charging rate of 40kW to 360kW, which means that to perform the same attack scenario we need 2100 EVs charging at 40 kW \textcolor{black}{or as little as 233 charging at 360kW superchargers}.

Now, oscillatory load attacks take advantage of load manipulation to impact the frequency stability of the power grid. This attack revolves around the concept of creating a demand surge to cause a frequency drop on the grid followed by a drop in demand to cause the frequency to overshoot. The adversary uses the compromised load to create an imbalance between the increased load and the generated power, causing the generators to slow down, hence resulting in a frequency drop. Consequently, the attacker switches off the compromised load to cause an increase in the frequency and the generator speed in response to the adversary's actions. The attacker alternates between charging and stopping or discharging to disturb and impact the grid. The variation of the power generation speed due to an oscillatory load attack with a 2.4 seconds period (1.2 seconds on and 1.2 seconds off) is demonstrated in Figure \ref{fig:speedAtt}. The sustained attack hinders the system's recovery causing fluctuations in the speed that would damage the turbines and decrease their lifetime due to the constant acceleration and deceleration. Different attacks could be launched by an adversary, however, a random delay between 0 to 4 seconds encompasses the wide range of attacks and is enough to mitigate this family of attacks. In future work, we will work on a mitigation mechanism that will utilize deep learning techniques to classify attacks based on their duty cycle that allows us to create a smart mitigation mechanism that minimizes the random delay needed.

\begin{figure}[t]
    \centering
    \includegraphics[width=0.74\linewidth]{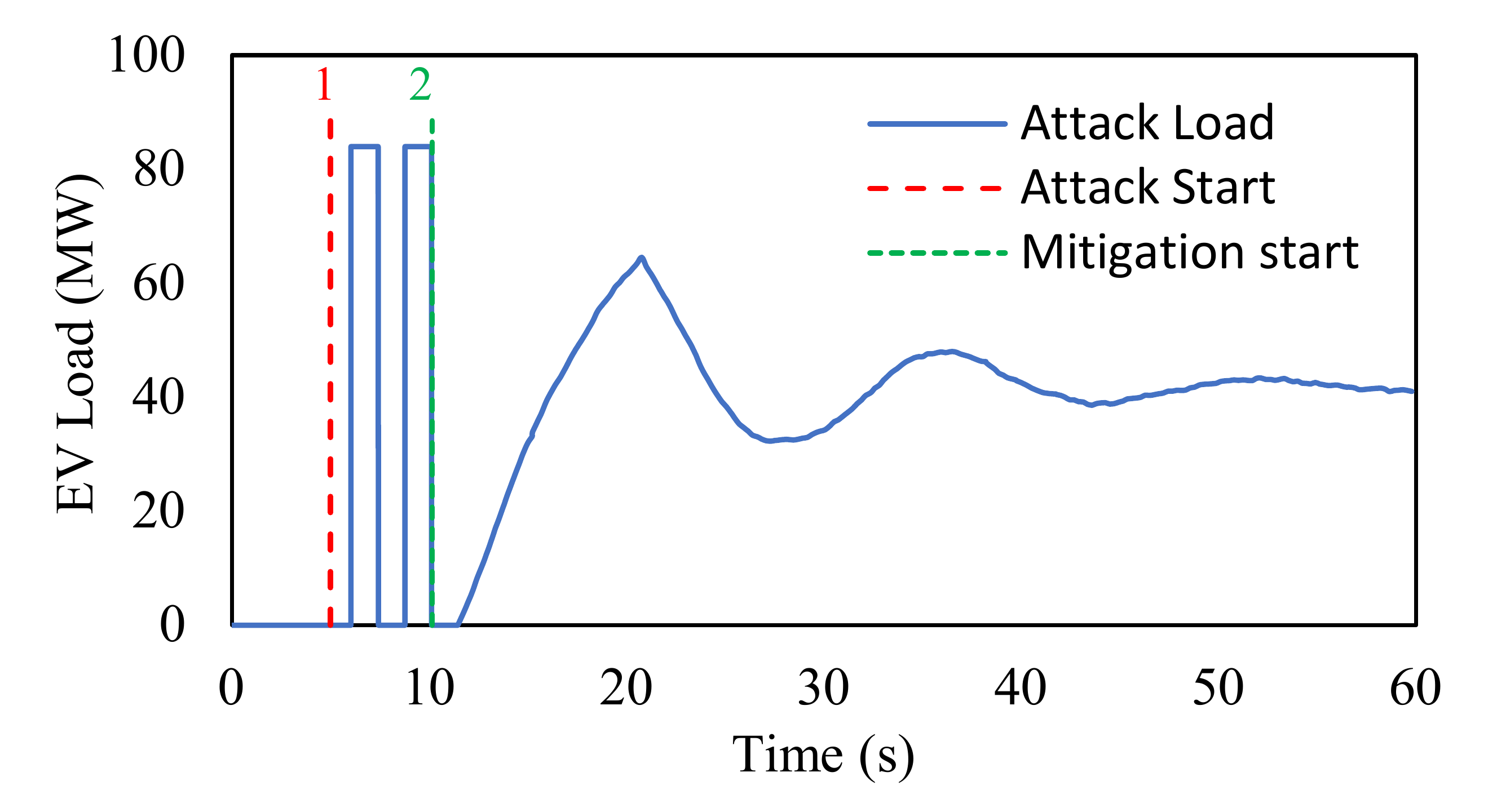}
    \caption{Load profile after mitigation. } 
    \label{fig:load}
\end{figure}

\begin{figure}[t]
     \centering
     \begin{subfigure}[b]{0.49\textwidth}
         \centering
         \includegraphics[width=\textwidth]{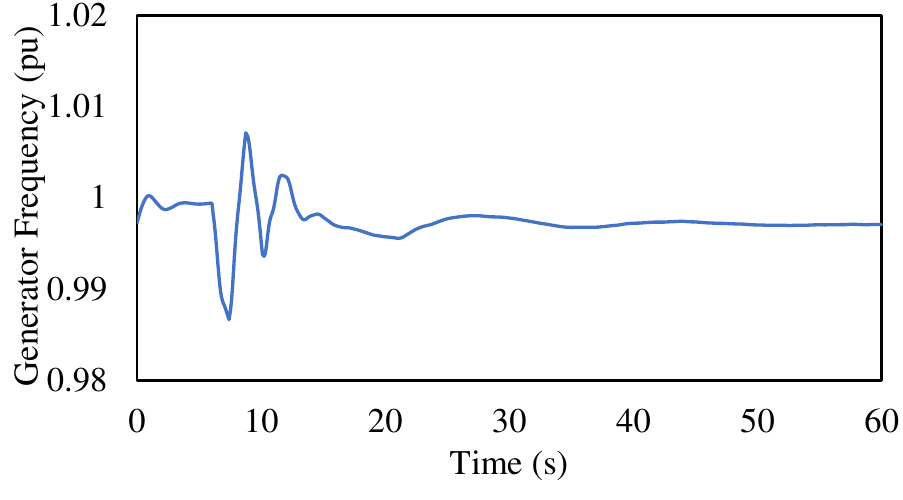}
         \caption{Generator 1}
     \end{subfigure}
     \hfill
     \begin{subfigure}[b]{0.49\textwidth}
         \centering
         \includegraphics[width=\textwidth]{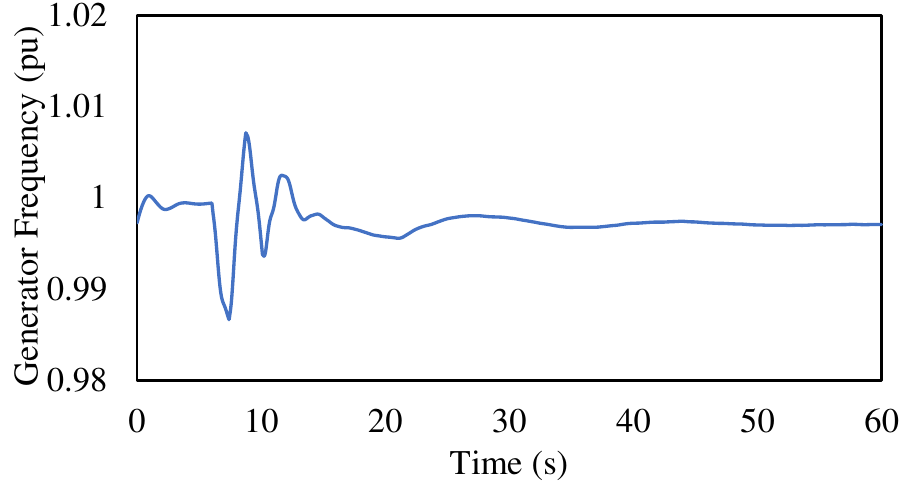}
         \caption{Generator 2}
     \end{subfigure}
     \hfill
    \begin{subfigure}[b]{0.49\textwidth}
         \centering
         \includegraphics[width=\textwidth]{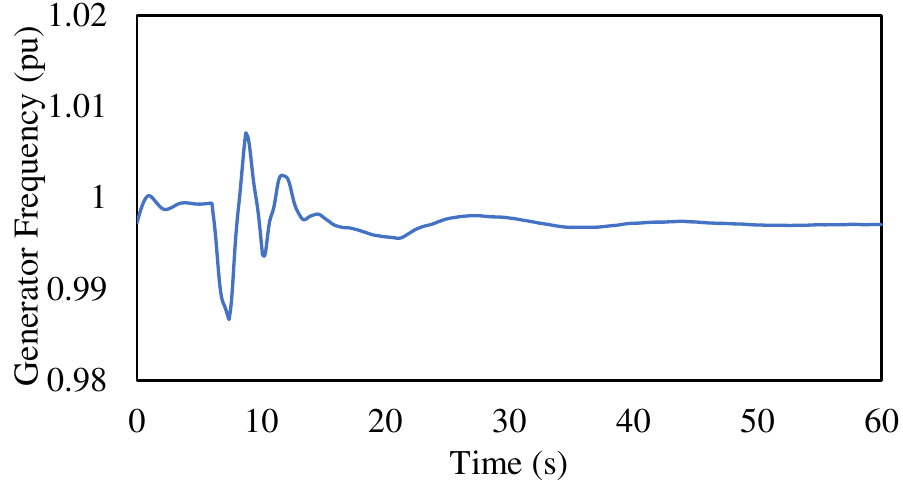}
         \caption{Generator 3}
    \end{subfigure}
    \caption{The variation of the generator's speed as a result of an oscillatory load attack followed by mitigation.} 
    \label{fig:speedMit}
\end{figure}

As shown in Figure \ref{fig:load}, the attack was launched (step 1) and detected (step 2) after 5 seconds using our novel detection mechanism and consequently, the different charging stations independently initiate their mitigation mechanism and induce a random delay between 0 and 4 seconds to attenuate the impact on the power grid and the generators and prevent attacker synchronization. The main aim here is to stabilize the system by ensuring the disturbance does not impact the system's performance. We demonstrate in Figure \ref{fig:load} the attack load and its variation over time after implementing our mitigation technique on our test bed (behavior of the grid after step 2). In the first few seconds, we observe two peaks showing the switching on and off due to an attack before detection, followed by a gradual distribution of the attack load over a period of 50 seconds as a result of the random delay of consequent start and stop requests. \textcolor{black}{It is clear in Figure\ref{fig:load} how the added random delay causes the discrete behavior of the attack load, especially beyond 40s}. The frequency variation after a mitigated attack is shown in Figure \ref{fig:speedMit}, that demonstrates the effectiveness of the proposed approach in eliminating the oscillations caused by the same sustained attack that is shown in Figure \ref{fig:speedAtt}.\textcolor{black}{To demonstrate the worst-case scenario that all the attacks are detected by the end of the first 5s of the attack and not before, we delayed the initiation of the mitigation technique till t=10s which is 5 seconds after the attack started}. This lightweight mitigation scheme prevents the attacker from synchronizing an attack thus, \textcolor{black}{eliminating the attacker's ability to impact the power grid. The randomization of the attack load over time results in a gradual increase in the load rather than instantaneous spikes and drops which allows the grid's generators to cope with the change in demand (behavior after step 2). Along with that, as shown in Figure \ref{fig:speedMit}, the generator speed starts to get damped at t=11s in 1s of detecting the attack. It is worth mentioning that stopping the attack doe not immediately bring the system back to its nominal speed. The system requires 2s to return from the abnormal state to the normal frequency range. Our mitigation strategy was successful in eliminating the impact of the attack on the power grid without the need for adding a physical control layer to the power grid. Furthermore, our proposed cyber-layer mitigation scheme is comparable to physical layer mitigation schemes. In \cite{kabir2021two}, the control scheme was able to limit the forced oscillation to a safe threshold after 15s of the attack initiation as compared to our 2s. Furthermore, our mitigation mechanism is able to completely eliminate the attack impact reducing the need for continuous acceleration and deceleration of the generators, unlike physical layer mitigation mechanisms that can dampen the dangerous oscillations but can never eliminate them.}

 Figure \ref{fig:load} demonstrates that after detection, the attack load would increase gradually to reach around 48 MW (half the total attack load). This would result in the grid's ability to return to stability since the attack load oscillations have been eliminated. It is worth mentioning that if the attack utilizes the V2G ability of the EVCSs, our mitigation would result in an EV attack load centered around zero, which is even better for the stability of the grid and allows the generator to regulate their speeds easily.


\section{Evaluation, Comparison, and Discussion}
\label{sec:results2}
To detect oscillatory load attacks, we proposed an approach that leverages the behavioral characteristics of the charging station and the power grid. \textcolor{black}{This approach is used to detect oscillatory load attacks initiated from the EV charging ecosystem regardless of the specific exploits and vulnerabilities used to compromise the different components that constitute it.} The events at the charging station are directly related to the behavior of the power grid, \textcolor{black}{making them the most important data on the cyber layer of the ecosystem}.

Furthermore, the result demonstrated that such coordinated attacks have a unique signature constituting of the charging event and the frequency on the power grid. Moreover, some attacks disguised as normal behavior (stealthy attacks) might go undetected. Based on our previous experiments, our choice of features significantly impacted our accuracy and allowed us to detect oscillatory load attacks with high recall and precision values. We evaluate our detection system on the 5-Attack dataset (5 seconds attack window) and the 10-Attack dataset (10 seconds attack windows). The precision and recall for LSTM and ConvLSTM improved as we increased the attack window from 5 seconds to 10 seconds, as summarized in Table \ref{table:outcomes}. The LSTM was less effective on the 5-Attack dataset and achieved an F-measure score of 97.493\%. However, the LSTM performance improved on the 10-Attack dataset. We observe improvement for other metrics studied (accuracy, precision, recall, and a number of false negative samples). The number of misclassified samples decreased from 40 to 10 when the LSTM model viewed a 10 seconds attack window, as shown in Table \ref{table:confusionMatrix}. \textcolor{black}{ Moreover, the 5-Attack LSTM model misclassified stealthy and non-stealthy attacks whereas, the ConvLSTM only misclassified few stealthy attacks achieving a low false negative on stealthy attacks with less than 1.7\% (0.88\% of total attacks). The same stealthy attacks were later detected in the 10-Attack ConvLSTM model decreasing the misclassified stealthy attacks to 0.19\% (0.099\% of total attacks) using the 10-Attack detector. Moreover, the other types of non-stealthy attacks were all detected using 5-Attack ConvLSTM. This shows that detecting stealthy oscillatory load attacks is not as trivial as other types of attacks and a two-step detection mechanism is best suited to provide redundancy and effectiveness to the detection mechanism to help secure the vital services provided by the power grid. It is worth noting that our detection mechanism works on a rolling basis as long as the charging station is receiving requests, thus, improving the detection ability of our proposed approach. We highlight that the rolling window 5-attack detector will detect attacks in any smaller time frame but we highlight that it was optimized for detection after 5 seconds to be able to detect stealthy attacks. We test our 5-Attack detector (without retaining) on windows containing only the first second of the attack behavior, and we were able to identify a third of the attacks within 1s of their initiation, achieving a recall of 33.3\%. In this test, only 3 normal data samples were misclassified as attacks which is consistent with the results of the original 5-Attack model. In the stealthy version of the attacks, the attacker launches slow oscillatory attacks as well as distributes the switching among multiple charging stations. This makes the behavior of individual EVCSs in short windows of time look normal since only one event or possibly no events at all occur. Indeed, this is aligned with our finding where the 5-attack model misclassified the stealthy attacks within the first second and labeled them as normal. This shows the need for a rolling window model. Consequently, we highlight that we chose the 5-Attack model so we do not compromise between the impact on the grid and the accuracy of our model. In a real-life deployment, the distributed mitigation mechanism in \ref{alg:two} will start mitigating the attacks as soon as the detector classifies a window as an anomaly and does not need to wait for 5 seconds after the attack is initiated}.

Moreover, the deep learning model ConvLSTM, an LSTM variant, achieved a better performance than the LSTM on the two datasets. The ConvLSTM showed improvement over LSTM on the 5-Attack dataset achieving a 99.405\% F-measure score. Moreover, the ConvLSTM also achieved a 99.803\% F-measure score on the 10-Attack dataset. It is important to note that the performance of the model 5-Attack dataset is crucial for our evaluation since early detection of the attack is needed. The LSTM did not perform as well on the 5-Attack dataset compared to the 10-Attack dataset, whereas the ConvLSTM using the convolutional filters on the input allowed the deep learning model to learn intricate patterns of the attack data which allowed it to effectively classify samples. \textcolor{black}{The ConvLSTM substitutes the matrix multiplication of the LSTM at each gate with convolutional operations that allowed it to extract the spatiotemporal relationships between the multiple timesteps over recorded variables (events and frequency). This relationship is the most crucial aspect in detecting oscillatory load attacks where ConvLSTM showed improved performance over the LSTM \cite{tariq2019detecting}. Moreover, the fully connected LSTM has to unfold the inputs to 1D vectors before processing them, thus losing all the spatial information during the process. Thus, to preserve the spatial features the ConvLSTM uses 3D tensors that preserve the spatial information and determine the future state of a cell by taking into consideration the local neighbors of a cell \cite{shi2015convolutional}. The ConvLSTM reaps the benefits of the LSTM with temporal data and the benefits of a convolutional neural network with spatial data, which was important in our study of the two features over multiple timesteps.} The ConvLSTM was able to learn the patterns of the attack with only 5-second windows. Indeed, the performance of both classifiers on the 10-Attack dataset is expected to be better, since the impact of the switching would increase tremendously showing a significant change in the behavior. Moreover, the number of misclassified samples, most importantly the false negatives, is crucial to evaluate the efficiency of our detection model and how much we can trust the classifiers to identify attacks. Our analysis presented in Table \ref{table:confusionMatrix} shows that the LSTM misclassified 40 and 10 attack samples as normal for the two datasets, which allows the adversary to execute a wider range of adversarial attacks as compared to the ConvLSTM which only misclassified 9 and 1 data samples for the 5-Attack and the 10 Attack datasets, respectively.

The ConvLSTM outperformed the LSTM in various aspects. However, training the ConvLSTM model took around two hours and a half as compared to the LSTM model which took about 4 minutes. The training time is tolerable in our system model because we assume that training is performed by a central authority with enough resources and has access to data from various operators \textcolor{black}{and does not impact the prediction of attacks. Moreover, the prediction time of the ConVLSTM is still in the order of milliseconds which means that although its complex structure requires extra training time, its performance once deployed is not hindered by this complexity. The complexity of the ConvLSTM model arises from the structure of the ConvLSTM layer that utilizes matrix multiplication along with the kernels and the filters used that increase the trainable parameters in the model drastically leading to high training time. Moreover, the increase in the training time of the ConvLSTM is also due to the batch size and the learning rate where increasing the batch size leads to a poor generalization over the data samples and a small learning rate requires more training epochs given the smaller changes made to the weights each update. In our approach a compromise was made between the training time and the accuracy to provide a reliable deep learning model that is able to effectively detect attacks}.  Moreover, the ConvLSTM and its convolutional mechanisms applied to features helped detect oscillatory switching attacks with as little as 5 seconds attack window. Further, to compare the computational performances of the devised classifiers, we measure their speed in terms of the time required to complete the classification experiments. As illustrated in Table \ref{table:time}, the LSTM performed significantly faster than the convolutional LSTM, with 0.003 seconds to complete the classification. Whereas ConvLSTM performed relatively slower, with a computational time of 0.011 seconds. However, the time required by the ConvLSTM is tolerable since the output of the deep learning is almost instantaneous.

Our deep learning model depends on the behavioral characteristics (logged by the charging station during operation). The characteristics allow us to distribute the decision-making where each charging station acts independently. To the best of our knowledge, we are the first to enable a distributed detection mechanism of oscillatory load attacks where models do not need to be deployed on a central management system to perform accurate detection. In \cite{kabir2021two}, Kabir et al. devised a detection algorithm that depends on two charging events and the number of vehicles connecting within $\Delta$ time. The number of vehicles is an artifact that is only known to the CMS of a specific operator and is not shared. However, in our approach, we utilize the frequency reading of the power grid, which is a shared variable (artifact) among the charging stations of different operators connected to the same bus. These features enable us to detect multi-operator and stealthy sophisticated attacks and distribute the detection mechanism. The deep learning model can be deployed on every charging station (public or private), where each EVCS can take its decision solely based on the artifacts (events and frequency) that can be collected by the EVCS independently. 

Furthermore, since our approach is deployed on the charging station itself it prevents MitM attacks that can be launched by an adversary on the OCPP traffic exchange between the CMS and the EVCS itself \cite{alcaraz2017ocpp} to control charging stations and perform oscillatory load attacks. Our detection approach mitigates various attack vectors by deploying the deep learning model on the component that is used to create an impact (EVCS). It is worth mentioning, that our detection mechanism could be deployed on private charging stations which mitigates the limitation introduced by central detection mechanisms \cite{kabir2021two}. Furthermore, our detection approach requires viewing only 5s of the oscillatory switching attacks as compared to \cite{kabir2021two} that was tested on 20, 30, and 40 seconds attack periods and resulted in 30\%, 10\%, 5\% false negative rates, respectively.

\textsc{Robustness and limitations}: In our approach, we address oscillatory load attacks and adversarial oscillatory attacks that other detection mechanisms fail to detect (e.g., multi-operator, stealthy, and MitM oscillatory load attacks). This improves the robustness of our model and increases the spectrum of various oscillatory load attacks that could be detected by this approach. Considering that our approach's scope is only to detect coordinated oscillatory load attacks based on the combination of cyber and physical behavioral characteristics, we performed our experiments on the New England 39-Bus System. We did not include data samples from different power grids. Attacks on the grids have various impacts. For example, an attack on a 9-Bus system might not have the same consequences on a 39-Bus system. However, our approach is easily reproducible to make it operational on other power grids. Although this work contributes to understanding and detecting oscillatory load attacks, however, it faces a few current limitations. For instance, the work relies on a supervised learning approach, which cannot classify new, previously unseen attacks. To overcome this limitation, unsupervised approaches can be considered complementary approaches to face the emergence of any adversarial attacks. Additionally, our approach can be leveraged as a stepping stone to develop new cyber-layer defense mechanisms that prevent and detect oscillatory load attacks. 

In our work, we assume that the charging station is honest, thus, the adversary can evade our detection mechanism by compromising the charging station itself. However, the adversary needs to compromise all the charging stations needed to mount attacks. The distributed nature of our detection mechanism makes it hard for the adversary to mount attacks easily and ensures fault tolerance in the system, unlike centralized detection mechanisms that provide a single point of failure. Moreover, adversaries need to hack and exploit charging stations of different firmware versions and would require the attacker to find vulnerabilities in the different types of charging stations. Finally, we plan in our future work to use federated learning to assist in preserving the privacy of the records during the training period without the need of the power grid operator to get charging behavior data to create an AI-enabled detection model.

\textcolor{black}{Moreover, it is important to note that to evade the mitigation technique, the attacker needs to guess the random number generator (if the operator/manufacturer used a weak random number generator). However, each charging station creates a random delay independently of the other hindering the adversary from discovering the random delay of all the charging stations that are being exploited to mount an oscillatory load attack.}
\textcolor{black}{Through this work, we show that a simple and lightweight random delay mechanism provides an efficient countermeasure to adversaries trying to launch oscillatory load attacks. This mechanism is compatible with the nature of the ecosystem as it doesn't require coordination with the other charging stations which would create an overhead for charging stations that are equipped with limited computing power. However, we plan in the future work to create a framework to support the grid using V2G and mitigate the impact of EVCS-launched cyber-attacks.}

\section{Conclusion}
\label{sec:conclusion}
The increase in the adoption of EVs and their IoT-enabled infrastructure is creating a new attack surface to target the power grid and cause instability to the infrastructure. \textcolor{black}{The different interconnected components can be exploited by} adversaries \textcolor{black}{to} initiate various types of oscillatory load attacks (e.g., stealthy, multi-operator, etc) \textcolor{black}{ that exploits a botnet of EVCSs (public or private) to induce and sustain grid instability}. Consequently, we devised a distributed deep learning \textcolor{black}{detection mechanism} that can accurately detect oscillatory load attacks by viewing as little as 5 seconds of an attack and achieving an F-measure score of 99.4\% with a false negative rate of less than 1\%. Our approach and the unique features we chose allowed us to deploy a reliable deep-learning model to detect attacks on public and private charging stations. Consequently, after detection, we evaluate the use of a \textcolor{black}{lightweight distributed mitigation mechanism that is also deployed on the charging station by }randomly delaying requests with a maximum delay of 4 seconds to ensure the quality of service in case of a false positive detection that would affect customers. We tested our distributed mitigation mechanism on a real-time EV co-simulation testbed. The detection/mitigation mechanisms proposed are robust, lightweight, and easily reproducible and provide a defense layer to secure the power grid. Moreover, this is a novel solution that can be deployed on existing infrastructure to build security into the ecosystem.

\section{Acknowledgement}

This research was conducted and funded as part of the Concordia University/ Hydro-Quebec/ NSERC research collaboration project ``Large-Scale Integration of EVCSs into the Smart Grid: A comprehensive cyber-physical study and security assessment." Grant reference: ALLRP 567144-21.

\bibliographystyle{elsarticle-num-names} 
\bibliography{main}

\begin{thebibliography}{60}
\expandafter\ifx\csname natexlab\endcsname\relax\def\natexlab#1{#1}\fi
\providecommand{\url}[1]{\texttt{#1}}
\providecommand{\href}[2]{#2}
\providecommand{\path}[1]{#1}
\providecommand{\DOIprefix}{doi:}
\providecommand{\ArXivprefix}{arXiv:}
\providecommand{\URLprefix}{URL: }
\providecommand{\Pubmedprefix}{pmid:}
\providecommand{\doi}[1]{\href{http://dx.doi.org/#1}{\path{#1}}}
\providecommand{\Pubmed}[1]{\href{pmid:#1}{\path{#1}}}
\providecommand{\bibinfo}[2]{#2}
\ifx\xfnm\relax \def\xfnm[#1]{\unskip,\space#1}\fi
\bibitem[{Regan(2020)}]{regan_2020}
\bibinfo{author}{Regan, H.},
\newblock \bibinfo{title}{China pledges to go carbon neutral by 2060},
\newblock \bibinfo{journal}{CNN}  (\bibinfo{year}{2020}). \URLprefix
  \url{https://www.cnn.com/2020/09/22/china/xi-jinping-carbon-neutral-2060-intl-hnk/index.html}.
\bibitem[{Flannery(2021)}]{flannery_2021}
\bibinfo{author}{Flannery, R.},
\newblock \bibinfo{title}{Ev share of china passenger car market more than
  tripled to nearly 19\% in october},
\newblock \bibinfo{journal}{Forbes}  (\bibinfo{year}{2021}). \URLprefix
  \url{https://www.forbes.com/sites/russellflannery/2021/11/13/ev-share-of-china-passenger-car-market-more-than-tripled-to-nearly-19-in-october/?sh=1591d22e2c21}.
\bibitem[{Shingler(2020)}]{shingler_2020}
\bibinfo{author}{Shingler, B.},
\newblock \bibinfo{title}{Quebec's push to go electric won't get province to
  emission reduction targets, experts say},
\newblock \bibinfo{journal}{CBCnews}  (\bibinfo{year}{2020}). \URLprefix
  \url{https://www.cbc.ca/news/canada/montreal/quebec-green-plan-1.5802976}.
\bibitem[{Gyulai(2020)}]{gyulai_2020}
\bibinfo{author}{Gyulai, L.},
\newblock \bibinfo{title}{Montreal's climate plan includes ban on non-electric
  cars downtown by 2030},
\newblock \bibinfo{journal}{Montreal Gazette}  (\bibinfo{year}{2020}).
  \URLprefix
  \url{https://montrealgazette.com/news/local-news/montreal-releases-climate-plan-including-ban-on-non-electric-cars-downtown-by-2030}.
\bibitem[{Riley(2021)}]{riley_2021}
\bibinfo{author}{Riley, C.},
\newblock \bibinfo{title}{Europe aims to kill gasoline and diesel cars by
  2035},
\newblock \bibinfo{journal}{CNN}  (\bibinfo{year}{2021}). \URLprefix
  \url{https://edition.cnn.com/2021/07/14/business/eu-emissions-climate-cars/index.html}.
\bibitem[{Khan et~al.(2022)Khan, El-Sayed, and Arboleya}]{khan2022multi}
\bibinfo{author}{Khan, K.}, \bibinfo{author}{El-Sayed, I.},
  \bibinfo{author}{Arboleya, P.},
\newblock \bibinfo{title}{Multi-issue negotiation evs charging mechanism in
  highly congested distribution networks},
\newblock \bibinfo{journal}{IEEE Transactions on Vehicular Technology}
  (\bibinfo{year}{2022}).
\bibitem[{Canada(2021)}]{canada_2021}
\bibinfo{author}{Canada, N.~R.},
\newblock \bibinfo{title}{Government of canada},
\newblock \bibinfo{journal}{Natural Resources Canada}  (\bibinfo{year}{2021}).
  \URLprefix
  \url{https://www.nrcan.gc.ca/energy-efficiency/transportation-alternative-fuels/zero-emission-vehicle-infrastructure-program/21876}.
\bibitem[{Sayed et~al.(2021)Sayed, Atallah, Assi, and
  Debbabi}]{sayed2021electric}
\bibinfo{author}{Sayed, M.~A.}, \bibinfo{author}{Atallah, R.},
  \bibinfo{author}{Assi, C.}, \bibinfo{author}{Debbabi, M.},
\newblock \bibinfo{title}{Electric vehicle attack impact on power grid
  operation},
\newblock \bibinfo{journal}{International Journal of Electrical Power \& Energy
  Systems}  (\bibinfo{year}{2021}) \bibinfo{pages}{107784}.
\bibitem[{Alcaraz et~al.(2017)Alcaraz, Lopez, and Wolthusen}]{alcaraz2017ocpp}
\bibinfo{author}{Alcaraz, C.}, \bibinfo{author}{Lopez, J.},
  \bibinfo{author}{Wolthusen, S.},
\newblock \bibinfo{title}{Ocpp protocol: Security threats and challenges},
\newblock \bibinfo{journal}{IEEE Transactions on Smart Grid}
  \bibinfo{volume}{8} (\bibinfo{year}{2017}) \bibinfo{pages}{2452--2459}.
\bibitem[{Rubio et~al.(2018)Rubio, Alcaraz, and Lopez}]{rubio2018addressing}
\bibinfo{author}{Rubio, J.~E.}, \bibinfo{author}{Alcaraz, C.},
  \bibinfo{author}{Lopez, J.},
\newblock \bibinfo{title}{Addressing security in ocpp: Protection against
  man-in-the-middle attacks},
\newblock in: \bibinfo{booktitle}{2018 9th IFIP International Conference on New
  Technologies, Mobility and Security (NTMS)}, \bibinfo{organization}{IEEE},
  \bibinfo{year}{2018}, pp. \bibinfo{pages}{1--5}.
\bibitem[{Massie(2022)}]{independent_2022}
\bibinfo{author}{Massie, G.},
\newblock \bibinfo{title}{Russian ev charging stations hacked.},
\newblock \bibinfo{journal}{The Independent}  (\bibinfo{year}{2022}).
  \URLprefix
  \url{https://www.independent.co.uk/news/world/europe/putin-charging-station-hacked-ukraine-russia-b2026260.html}.
\bibitem[{Thornburn(2022)}]{mailonline_2022}
\bibinfo{author}{Thornburn, J.}, \bibinfo{title}{Isle of wight: Ev charge
  points are hacked to show porn websites}, \bibinfo{year}{2022}. \URLprefix
  \url{https://www.dailymail.co.uk/news/article-10692981/Isle-Wight-EV-charge-points-hacked-PORN-websites.html}.
\bibitem[{Polityuk et~al.(2017)Polityuk, Vukmanovic, and
  Jewkes}]{polityuk_vukmanovic_jewkes_2017}
\bibinfo{author}{Polityuk, P.}, \bibinfo{author}{Vukmanovic, O.},
  \bibinfo{author}{Jewkes, S.},
\newblock \bibinfo{title}{Ukraine's power outage was a cyber attack:
  Ukrenergo},
\newblock \bibinfo{journal}{Reuters}  (\bibinfo{year}{2017}). \URLprefix
  \url{https://www.reuters.com/article/us-ukraine-cyber-attack-energy-idUSKBN1521BA}.
\bibitem[{Perez(2016)}]{perez_2016}
\bibinfo{author}{Perez, E.},
\newblock \bibinfo{title}{First on cnn: U.s. investigators find proof of
  cyberattack on ukraine power grid | cnn politics},
\newblock \bibinfo{journal}{CNN}  (\bibinfo{year}{2016}). \URLprefix
  \url{https://www.cnn.com/2016/02/03/politics/cyberattack-ukraine-power-grid/index.html}.
\bibitem[{Soltan et~al.(2018)Soltan, Mittal, and Poor}]{soltan2018blackiot}
\bibinfo{author}{Soltan, S.}, \bibinfo{author}{Mittal, P.},
  \bibinfo{author}{Poor, H.~V.},
\newblock \bibinfo{title}{Blackiot: Iot botnet of high wattage devices can
  disrupt the power grid},
\newblock in: \bibinfo{booktitle}{27th $\{$USENIX$\}$ Security Symposium
  ($\{$USENIX$\}$ Security 18)}, \bibinfo{year}{2018}, pp.
  \bibinfo{pages}{15--32}.
\bibitem[{Sarieddine et~al.(2022)Sarieddine, Sayed, Torabi, Atallah, and
  Assi}]{sarieddine2022investigating}
\bibinfo{author}{Sarieddine, K.}, \bibinfo{author}{Sayed, M.-A.},
  \bibinfo{author}{Torabi, S.}, \bibinfo{author}{Atallah, R.},
  \bibinfo{author}{Assi, C.},
\newblock \bibinfo{title}{Investigating the security of ev charging mobile
  applications as an attack surface},
\newblock \bibinfo{journal}{arXiv preprint arXiv:2211.10603}
  (\bibinfo{year}{2022}).
\bibitem[{Akhras et~al.(2020)Akhras, El-Hajj, Majdalani, Hajj, Jabr, and
  Shaban}]{akhras2020securing}
\bibinfo{author}{Akhras, R.}, \bibinfo{author}{El-Hajj, W.},
  \bibinfo{author}{Majdalani, M.}, \bibinfo{author}{Hajj, H.},
  \bibinfo{author}{Jabr, R.}, \bibinfo{author}{Shaban, K.},
\newblock \bibinfo{title}{Securing smart grid communication using ethereum
  smart contracts},
\newblock in: \bibinfo{booktitle}{2020 International Wireless Communications
  and Mobile Computing (IWCMC)}, \bibinfo{organization}{IEEE},
  \bibinfo{year}{2020}, pp. \bibinfo{pages}{1672--1678}.
\bibitem[{Upadhyay et~al.(2021)Upadhyay, Manero, Zaman, and
  Sampalli}]{upadhyay2021intrusion}
\bibinfo{author}{Upadhyay, D.}, \bibinfo{author}{Manero, J.},
  \bibinfo{author}{Zaman, M.}, \bibinfo{author}{Sampalli, S.},
\newblock \bibinfo{title}{Intrusion detection in scada based power grids:
  Recursive feature elimination model with majority vote ensemble algorithm},
\newblock \bibinfo{journal}{IEEE Transactions on Network Science and
  Engineering} \bibinfo{volume}{8} (\bibinfo{year}{2021})
  \bibinfo{pages}{2559--2574}.
\bibitem[{Upadhyay et~al.(2020)Upadhyay, Manero, Zaman, and
  Sampalli}]{upadhyay2020gradient}
\bibinfo{author}{Upadhyay, D.}, \bibinfo{author}{Manero, J.},
  \bibinfo{author}{Zaman, M.}, \bibinfo{author}{Sampalli, S.},
\newblock \bibinfo{title}{Gradient boosting feature selection with machine
  learning classifiers for intrusion detection on power grids},
\newblock \bibinfo{journal}{IEEE Transactions on Network and Service
  Management} \bibinfo{volume}{18} (\bibinfo{year}{2020})
  \bibinfo{pages}{1104--1116}.
\bibitem[{Li et~al.(2021)Li, Xue, Wu, Wang, Zhou, Aziz, and
  He}]{li2021intrusion}
\bibinfo{author}{Li, Y.}, \bibinfo{author}{Xue, W.}, \bibinfo{author}{Wu, T.},
  \bibinfo{author}{Wang, H.}, \bibinfo{author}{Zhou, B.},
  \bibinfo{author}{Aziz, S.}, \bibinfo{author}{He, Y.},
\newblock \bibinfo{title}{Intrusion detection of cyber physical energy system
  based on multivariate ensemble classification},
\newblock \bibinfo{journal}{Energy} \bibinfo{volume}{218}
  (\bibinfo{year}{2021}) \bibinfo{pages}{119505}.
\bibitem[{Kabir et~al.(2021)Kabir, Ghafouri, Moussa, and Assi}]{kabir2021two}
\bibinfo{author}{Kabir, M.~E.}, \bibinfo{author}{Ghafouri, M.},
  \bibinfo{author}{Moussa, B.}, \bibinfo{author}{Assi, C.},
\newblock \bibinfo{title}{A two-stage protection method for detection and
  mitigation of coordinated evse switching attacks},
\newblock \bibinfo{journal}{IEEE Transactions on Smart Grid}
  (\bibinfo{year}{2021}).
\bibitem[{Alliance(2021)}]{ocpp}
\bibinfo{author}{Alliance, O.~C.},
\newblock \bibinfo{title}{Open charge point protocol 2.0.1},
\newblock \bibinfo{journal}{Open Charge Alliance}  (\bibinfo{year}{2021}).
  \URLprefix \url{https://www.openchargealliance.org/protocols/ocpp-201/}.
\bibitem[{Dmitry(2018)}]{kaspersky}
\bibinfo{author}{Dmitry, S.},
\newblock \bibinfo{title}{{ChargePoint Home Security Research}}
  (\bibinfo{year}{2018}).
\bibitem[{Ayad et~al.(2018)Ayad, Farag, Youssef, and
  El-Saadany}]{ayad2018detection}
\bibinfo{author}{Ayad, A.}, \bibinfo{author}{Farag, H.~E.},
  \bibinfo{author}{Youssef, A.}, \bibinfo{author}{El-Saadany, E.~F.},
\newblock \bibinfo{title}{Detection of false data injection attacks in smart
  grids using recurrent neural networks},
\newblock in: \bibinfo{booktitle}{2018 IEEE Power \& Energy Society Innovative
  Smart Grid Technologies Conference (ISGT)}, \bibinfo{organization}{IEEE},
  \bibinfo{year}{2018}, pp. \bibinfo{pages}{1--5}.
\bibitem[{Margossian et~al.(2019)Margossian, Sayed, Fawaz, and
  Nakad}]{margossian2019partial}
\bibinfo{author}{Margossian, H.}, \bibinfo{author}{Sayed, M.~A.},
  \bibinfo{author}{Fawaz, W.}, \bibinfo{author}{Nakad, Z.},
\newblock \bibinfo{title}{Partial grid false data injection attacks against
  state estimation},
\newblock \bibinfo{journal}{International Journal of Electrical Power \& Energy
  Systems} \bibinfo{volume}{110} (\bibinfo{year}{2019})
  \bibinfo{pages}{623--629}.
\bibitem[{Kosut et~al.(2011)Kosut, Jia, Thomas, and Tong}]{kosut2011malicious}
\bibinfo{author}{Kosut, O.}, \bibinfo{author}{Jia, L.},
  \bibinfo{author}{Thomas, R.~J.}, \bibinfo{author}{Tong, L.},
\newblock \bibinfo{title}{Malicious data attacks on the smart grid},
\newblock \bibinfo{journal}{IEEE Transactions on Smart Grid}
  \bibinfo{volume}{2} (\bibinfo{year}{2011}) \bibinfo{pages}{645--658}.
\bibitem[{Li et~al.(2014)Li, Y{\i}lmaz, and Wang}]{li2014quickest}
\bibinfo{author}{Li, S.}, \bibinfo{author}{Y{\i}lmaz, Y.},
  \bibinfo{author}{Wang, X.},
\newblock \bibinfo{title}{Quickest detection of false data injection attack in
  wide-area smart grids},
\newblock \bibinfo{journal}{IEEE Transactions on Smart Grid}
  \bibinfo{volume}{6} (\bibinfo{year}{2014}) \bibinfo{pages}{2725--2735}.
\bibitem[{Wang et~al.(2021)Wang, Zhang, Ma, and Jin}]{wang2021kfrnn}
\bibinfo{author}{Wang, Y.}, \bibinfo{author}{Zhang, Z.}, \bibinfo{author}{Ma,
  J.}, \bibinfo{author}{Jin, Q.},
\newblock \bibinfo{title}{Kfrnn: An effective false data injection attack
  detection in smart grid based on kalman filter and recurrent neural network},
\newblock \bibinfo{journal}{IEEE Internet of Things Journal}
  (\bibinfo{year}{2021}).
\bibitem[{Sakhnini et~al.(2021)Sakhnini, Karimipour, Dehghantanha, and
  Parizi}]{sakhnini2021physical}
\bibinfo{author}{Sakhnini, J.}, \bibinfo{author}{Karimipour, H.},
  \bibinfo{author}{Dehghantanha, A.}, \bibinfo{author}{Parizi, R.~M.},
\newblock \bibinfo{title}{Physical layer attack identification and localization
  in cyber--physical grid: An ensemble deep learning based approach},
\newblock \bibinfo{journal}{Physical Communication} \bibinfo{volume}{47}
  (\bibinfo{year}{2021}) \bibinfo{pages}{101394}.
\bibitem[{Chung et~al.(2018)Chung, Li, Yuen, Chung, Zhang, and
  Wen}]{chung2018local}
\bibinfo{author}{Chung, H.-M.}, \bibinfo{author}{Li, W.-T.},
  \bibinfo{author}{Yuen, C.}, \bibinfo{author}{Chung, W.-H.},
  \bibinfo{author}{Zhang, Y.}, \bibinfo{author}{Wen, C.-K.},
\newblock \bibinfo{title}{Local cyber-physical attack for masking line outage
  and topology attack in smart grid},
\newblock \bibinfo{journal}{IEEE Transactions on Smart Grid}
  \bibinfo{volume}{10} (\bibinfo{year}{2018}) \bibinfo{pages}{4577--4588}.
\bibitem[{Basnet and Ali(2020)}]{basnet2020deep}
\bibinfo{author}{Basnet, M.}, \bibinfo{author}{Ali, M.~H.},
\newblock \bibinfo{title}{Deep learning-based intrusion detection system for
  electric vehicle charging station},
\newblock in: \bibinfo{booktitle}{2020 2nd International Conference on Smart
  Power \& Internet Energy Systems (SPIES)}, \bibinfo{organization}{IEEE},
  \bibinfo{year}{2020}, pp. \bibinfo{pages}{408--413}.
\bibitem[{Basnet and Ali(2021)}]{basnet2021exploring}
\bibinfo{author}{Basnet, M.}, \bibinfo{author}{Ali, M.~H.},
\newblock \bibinfo{title}{Exploring cybersecurity issues in 5g enabled electric
  vehicle charging station with deep learning},
\newblock \bibinfo{journal}{arXiv preprint arXiv:2104.08553}
  (\bibinfo{year}{2021}).
\bibitem[{Basnet et~al.(2021)Basnet, Poudyal, Ali, and
  Dasgupta}]{basnet2021ransomware}
\bibinfo{author}{Basnet, M.}, \bibinfo{author}{Poudyal, S.},
  \bibinfo{author}{Ali, M.~H.}, \bibinfo{author}{Dasgupta, D.},
\newblock \bibinfo{title}{Ransomware detection using deep learning in the scada
  system of electric vehicle charging station},
\newblock in: \bibinfo{booktitle}{2021 IEEE PES Innovative Smart Grid
  Technologies Conference-Latin America (ISGT Latin America)},
  \bibinfo{organization}{IEEE}, \bibinfo{year}{2021}, pp.
  \bibinfo{pages}{1--5}.
\bibitem[{Molina et~al.(2021)Molina, Torabi, Sarieddine, Bou-Harb, Bouguila,
  and Assi}]{sarieddine2021ransomware}
\bibinfo{author}{Molina, R. M.~A.}, \bibinfo{author}{Torabi, S.},
  \bibinfo{author}{Sarieddine, K.}, \bibinfo{author}{Bou-Harb, E.},
  \bibinfo{author}{Bouguila, N.}, \bibinfo{author}{Assi, C.},
\newblock \bibinfo{title}{On ransomware family attribution using pre-attack
  paranoia activities},
\newblock \bibinfo{journal}{IEEE Transactions on Network and Service
  Management}  (\bibinfo{year}{2021}).
\bibitem[{Acharya et~al.(2020)Acharya, Dvorkin, and Karri}]{acharya2020public}
\bibinfo{author}{Acharya, S.}, \bibinfo{author}{Dvorkin, Y.},
  \bibinfo{author}{Karri, R.},
\newblock \bibinfo{title}{Public plug-in electric vehicles+ grid data: Is a new
  cyberattack vector viable?},
\newblock \bibinfo{journal}{IEEE Transactions on Smart Grid}
  \bibinfo{volume}{11} (\bibinfo{year}{2020}) \bibinfo{pages}{5099--5113}.
\bibitem[{Hammad et~al.(2017)Hammad, Khalil, Farraj, Kundur, and
  Iravani}]{hammad2017class}
\bibinfo{author}{Hammad, E.}, \bibinfo{author}{Khalil, A.~M.},
  \bibinfo{author}{Farraj, A.}, \bibinfo{author}{Kundur, D.},
  \bibinfo{author}{Iravani, R.},
\newblock \bibinfo{title}{A class of switching exploits based on inter-area
  oscillations},
\newblock \bibinfo{journal}{IEEE Transactions on Smart Grid}
  \bibinfo{volume}{9} (\bibinfo{year}{2017}) \bibinfo{pages}{4659--4668}.
\bibitem[{Ghafouri et~al.(2022)Ghafouri, Kabir, Moussa, and
  Assi}]{ghafouri2022coordinated}
\bibinfo{author}{Ghafouri, M.}, \bibinfo{author}{Kabir, M.~E.},
  \bibinfo{author}{Moussa, B.}, \bibinfo{author}{Assi, C.},
\newblock \bibinfo{title}{Coordinated charging and discharging of electric
  vehicles: A new class of switching attacks},
\newblock \bibinfo{journal}{ACM Transactions on Cyber-Physical Systems}
  (\bibinfo{year}{2022}).
\bibitem[{Sayed et~al.(2022)Sayed, Ghafouri, Debbabi, and
  Assi}]{sayed2022dynamic}
\bibinfo{author}{Sayed, M.~A.}, \bibinfo{author}{Ghafouri, M.},
  \bibinfo{author}{Debbabi, M.}, \bibinfo{author}{Assi, C.},
\newblock \bibinfo{title}{Dynamic load altering ev attacks against power grid
  frequency control},
\newblock in: \bibinfo{booktitle}{2022 IEEE Power \& Energy Society General
  Meeting (PESGM)}, \bibinfo{organization}{IEEE}, \bibinfo{year}{2022}, pp.
  \bibinfo{pages}{1--5}.
\bibitem[{Hodge et~al.(2019)Hodge, Hauck, Gupta, and
  Bennett}]{hodge2019vehicle}
\bibinfo{author}{Hodge, C.}, \bibinfo{author}{Hauck, K.},
  \bibinfo{author}{Gupta, S.}, \bibinfo{author}{Bennett, J.~C.},
  \bibinfo{title}{Vehicle cybersecurity threats and mitigation approaches},
  \bibinfo{type}{Technical Report}, National Renewable Energy Lab.(NREL),
  Golden, CO (United States), \bibinfo{year}{2019}.
\bibitem[{Du et~al.(2021)Du, Yan, Ghafouri, Zgheib, Kassouf, and
  Debbabi}]{du2021modeling}
\bibinfo{author}{Du, H.}, \bibinfo{author}{Yan, J.}, \bibinfo{author}{Ghafouri,
  M.}, \bibinfo{author}{Zgheib, R.}, \bibinfo{author}{Kassouf, M.},
  \bibinfo{author}{Debbabi, M.},
\newblock \bibinfo{title}{Modeling of cyber attacks against converter-driven
  stability of pmsg-based wind farms with intentional subsynchronous
  resonance},
\newblock in: \bibinfo{booktitle}{2021 IEEE International Conference on
  Communications, Control, and Computing Technologies for Smart Grids
  (SmartGridComm)}, \bibinfo{organization}{IEEE}, \bibinfo{year}{2021}, pp.
  \bibinfo{pages}{391--397}.
\bibitem[{Acharya et~al.(2020)Acharya, Dvorkin, Pand{\v{z}}i{\'c}, and
  Karri}]{acharya2020cybersecurity}
\bibinfo{author}{Acharya, S.}, \bibinfo{author}{Dvorkin, Y.},
  \bibinfo{author}{Pand{\v{z}}i{\'c}, H.}, \bibinfo{author}{Karri, R.},
\newblock \bibinfo{title}{Cybersecurity of smart electric vehicle charging: A
  power grid perspective},
\newblock \bibinfo{journal}{IEEE Access} \bibinfo{volume}{8}
  (\bibinfo{year}{2020}) \bibinfo{pages}{214434--214453}.
\bibitem[{Lucia and Youssef(2021)}]{lucia2021covert}
\bibinfo{author}{Lucia, W.}, \bibinfo{author}{Youssef, A.},
\newblock \bibinfo{title}{Covert channels in stochastic cyber-physical
  systems},
\newblock \bibinfo{journal}{IET Cyber-Physical Systems: Theory \& Applications}
  \bibinfo{volume}{6} (\bibinfo{year}{2021}) \bibinfo{pages}{228--237}.
\bibitem[{Maddox(2010)}]{maddox_2010}
\bibinfo{author}{Maddox, J.},
\newblock \bibinfo{title}{Stuxnet: Malware more complex, targeted and dangerous
  than ever},
\newblock \bibinfo{journal}{CNN}  (\bibinfo{year}{2010}). \URLprefix
  \url{http://www.cnn.com/2010/TECH/innovation/09/24/stuxnet.computer.malware/index.html}.
\bibitem[{aem(2022)}]{aemo_2022}
\bibinfo{title}{Australian energy market operator},
\newblock \bibinfo{journal}{AEMO}  (\bibinfo{year}{2022}). \URLprefix
  \url{https://aemo.com.au/en}.
\bibitem[{aus(????)}]{australianas}
\bibinfo{title}{Motor vehicle census, australia},
\newblock \bibinfo{journal}{Australian Bureau of Statistics}  (????).
  \URLprefix
  \url{https://www.abs.gov.au/statistics/industry/tourism-and-transport/motor-vehicle-census-australia}.
\bibitem[{IEA(2022)}]{iea}
\bibinfo{author}{IEA},
\newblock \bibinfo{title}{Global ev outlook 2022 – analysis},
\newblock \bibinfo{journal}{IEA}  (\bibinfo{year}{2022}). \URLprefix
  \url{https://www.iea.org/reports/global-ev-outlook-2022}.
\bibitem[{Khoury and Nassar(2020)}]{khoury2020hybrid}
\bibinfo{author}{Khoury, J.}, \bibinfo{author}{Nassar, M.},
\newblock \bibinfo{title}{A hybrid game theory and reinforcement learning
  approach for cyber-physical systems security},
\newblock in: \bibinfo{booktitle}{NOMS 2020-2020 IEEE/IFIP Network Operations
  and Management Symposium}, \bibinfo{organization}{IEEE},
  \bibinfo{year}{2020}, pp. \bibinfo{pages}{1--9}.
\bibitem[{Athay et~al.(1979)Athay, Podmore, and Virmani}]{athay1979practical}
\bibinfo{author}{Athay, T.}, \bibinfo{author}{Podmore, R.},
  \bibinfo{author}{Virmani, S.},
\newblock \bibinfo{title}{A practical method for the direct analysis of
  transient stability},
\newblock \bibinfo{journal}{IEEE Transactions on Power Apparatus and Systems}
  (\bibinfo{year}{1979}) \bibinfo{pages}{573--584}.
\bibitem[{Bhatt and Affljulla(2017)}]{bhatt2017analysis}
\bibinfo{author}{Bhatt, G.}, \bibinfo{author}{Affljulla, S.},
\newblock \bibinfo{title}{Analysis of large scale pv penetration impact on ieee
  39-bus power system},
\newblock in: \bibinfo{booktitle}{2017 IEEE 58th International Scientific
  Conference on Power and Electrical Engineering of Riga Technical University
  (RTUCON)}, \bibinfo{organization}{IEEE}, \bibinfo{year}{2017}, pp.
  \bibinfo{pages}{1--6}.
\bibitem[{Shi et~al.(2015)Shi, Chen, Wang, Yeung, Wong, and
  Woo}]{shi2015convolutional}
\bibinfo{author}{Shi, X.}, \bibinfo{author}{Chen, Z.}, \bibinfo{author}{Wang,
  H.}, \bibinfo{author}{Yeung, D.-Y.}, \bibinfo{author}{Wong, W.-K.},
  \bibinfo{author}{Woo, W.-c.},
\newblock \bibinfo{title}{Convolutional lstm network: A machine learning
  approach for precipitation nowcasting},
\newblock \bibinfo{journal}{Advances in neural information processing systems}
  \bibinfo{volume}{28} (\bibinfo{year}{2015}).
\bibitem[{Zhang et~al.(2022)Zhang, Shi, Zhang, Cao, and
  Terzija}]{zhang2022review}
\bibinfo{author}{Zhang, Y.}, \bibinfo{author}{Shi, X.}, \bibinfo{author}{Zhang,
  H.}, \bibinfo{author}{Cao, Y.}, \bibinfo{author}{Terzija, V.},
\newblock \bibinfo{title}{Review on deep learning applications in frequency
  analysis and control of modern power system},
\newblock \bibinfo{journal}{International Journal of Electrical Power \& Energy
  Systems} \bibinfo{volume}{136} (\bibinfo{year}{2022})
  \bibinfo{pages}{107744}.
\bibitem[{Marshall et~al.(2018)Marshall, Brown, Fritz, and
  Johnson}]{marshall2018mastering}
\bibinfo{author}{Marshall, N.}, \bibinfo{author}{Brown, M.},
  \bibinfo{author}{Fritz, G.~B.}, \bibinfo{author}{Johnson, R.},
  \bibinfo{title}{Mastering VMware VSphere 6.7}, \bibinfo{publisher}{John Wiley
  \& Sons}, \bibinfo{year}{2018}.
\bibitem[{Team(2022)}]{adamoptimizer}
\bibinfo{author}{Team, K.},
\newblock \bibinfo{title}{Keras documentation: Adam},
\newblock \bibinfo{journal}{Keras}  (\bibinfo{year}{2022}). \URLprefix
  \url{https://keras.io/api/optimizers/adam/}.
\bibitem[{Ruder(2016)}]{ruder2016overview}
\bibinfo{author}{Ruder, S.},
\newblock \bibinfo{title}{An overview of gradient descent optimization
  algorithms},
\newblock \bibinfo{journal}{arXiv preprint arXiv:1609.04747}
  (\bibinfo{year}{2016}).
\bibitem[{Srivastava et~al.(2014)Srivastava, Hinton, Krizhevsky, Sutskever, and
  Salakhutdinov}]{srivastava2014dropout}
\bibinfo{author}{Srivastava, N.}, \bibinfo{author}{Hinton, G.},
  \bibinfo{author}{Krizhevsky, A.}, \bibinfo{author}{Sutskever, I.},
  \bibinfo{author}{Salakhutdinov, R.},
\newblock \bibinfo{title}{Dropout: a simple way to prevent neural networks from
  overfitting},
\newblock \bibinfo{journal}{The journal of machine learning research}
  \bibinfo{volume}{15} (\bibinfo{year}{2014}) \bibinfo{pages}{1929--1958}.
\bibitem[{Santurkar et~al.(2018)Santurkar, Tsipras, Ilyas, and
  Madry}]{santurkar2018does}
\bibinfo{author}{Santurkar, S.}, \bibinfo{author}{Tsipras, D.},
  \bibinfo{author}{Ilyas, A.}, \bibinfo{author}{Madry, A.},
\newblock \bibinfo{title}{How does batch normalization help optimization?},
\newblock \bibinfo{journal}{Advances in neural information processing systems}
  \bibinfo{volume}{31} (\bibinfo{year}{2018}).
\bibitem[{Mantovani et~al.(2015)Mantovani, Rossi, Vanschoren, Bischl, and
  De~Carvalho}]{mantovani2015effectiveness}
\bibinfo{author}{Mantovani, R.~G.}, \bibinfo{author}{Rossi, A.~L.},
  \bibinfo{author}{Vanschoren, J.}, \bibinfo{author}{Bischl, B.},
  \bibinfo{author}{De~Carvalho, A.~C.},
\newblock \bibinfo{title}{Effectiveness of random search in svm hyper-parameter
  tuning},
\newblock in: \bibinfo{booktitle}{2015 International Joint Conference on Neural
  Networks (IJCNN)}, \bibinfo{organization}{Ieee}, \bibinfo{year}{2015}, pp.
  \bibinfo{pages}{1--8}.
\bibitem[{Institute(2022)}]{illinois}
\bibinfo{author}{Institute, I.~T.},
\newblock \bibinfo{title}{Illinois center for a smarter electric grid (icseg)},
\newblock \bibinfo{journal}{Illinois Center for a Smarter Electric Grid ICSEG
  WSCC 9Bus System Comments}  (\bibinfo{year}{2022}). \URLprefix
  \url{https://icseg.iti.illinois.edu/wscc-9-bus-system/}.
\bibitem[{Galletta et~al.(2004)Galletta, Henry, McCoy, and
  Polak}]{galletta2004web}
\bibinfo{author}{Galletta, D.~F.}, \bibinfo{author}{Henry, R.},
  \bibinfo{author}{McCoy, S.}, \bibinfo{author}{Polak, P.},
\newblock \bibinfo{title}{Web site delays: How tolerant are users?},
\newblock \bibinfo{journal}{Journal of the Association for Information Systems}
  \bibinfo{volume}{5} (\bibinfo{year}{2004}) \bibinfo{pages}{1}.
\bibitem[{Tariq et~al.(2019)Tariq, Lee, Shin, Lee, Jung, Chung, and
  Woo}]{tariq2019detecting}
\bibinfo{author}{Tariq, S.}, \bibinfo{author}{Lee, S.}, \bibinfo{author}{Shin,
  Y.}, \bibinfo{author}{Lee, M.~S.}, \bibinfo{author}{Jung, O.},
  \bibinfo{author}{Chung, D.}, \bibinfo{author}{Woo, S.~S.},
\newblock \bibinfo{title}{Detecting anomalies in space using multivariate
  convolutional lstm with mixtures of probabilistic pca},
\newblock in: \bibinfo{booktitle}{Proceedings of the 25th ACM SIGKDD
  international conference on knowledge discovery \& data mining},
  \bibinfo{year}{2019}, pp. \bibinfo{pages}{2123--2133}.

\end{thebibliography}





\end{document}